\documentclass[12pt]{article}
\usepackage{amssymb}
\usepackage{bbm}
\usepackage{epsfig}
\usepackage{array}
\usepackage{float}
\usepackage{dsfont}
\usepackage{amstext}
\usepackage{amsmath}
\usepackage{graphicx}
\usepackage{mathabx}
\usepackage{psfrag}
\usepackage{a4}
\usepackage{a4wide}
\usepackage{xcolor}
\definecolor{darkteal}{RGB}{0,125,114}
\usepackage[colorlinks=true
,urlcolor=darkteal
,anchorcolor=darkteal
,citecolor=darkteal
,filecolor=darkteal
,linkcolor=darkteal
,menucolor=darkteal
,pagecolor=darkteal
,linktocpage=true
,pdfproducer=medialab
,pdfa=true
]{hyperref}
\allowdisplaybreaks[1]
\usepackage{multicol} 
\usepackage{multirow}
\usepackage{subcaption}
\usepackage{ulem}
\usepackage {setspace}
\usepackage{xcolor,colortbl}
\def\be{\begin{equation}}
	\def\ee{\end{equation}}
\def\bea{\begin{eqnarray}}
	\def\eea{\end{eqnarray}}
\DeclareMathOperator{\Tr}{Tr}

\newcolumntype{g}{>{\columncolor{Gray}}c}
\newcolumntype{w}{>{\columncolor{Gray2}}c}
\definecolor{Gray}{gray}{0.95}
\definecolor{Gray2}{gray}{0.98}
\begin{document}
	
	\begin{center}
		\baselineskip 20pt 
		{\large\bf Probing Stochastic Gravitational Wave Background from $SU(5) \times U(1)_{\chi}$ Strings in Light of NANOGrav 15-Year Data}
		\vspace{1cm}
		
		{\textbf{Waqas Ahmed,}$^{a}$\footnote{E-mail: \texttt{\href{mailto: waqasmit@hbpu.edu.cn}{waqasmit@hbpu.edu.cn}}}
			\textbf{Mansoor Ur Rehman}$^{b}$\footnote{E-mail: \texttt{\href{mailto: mansoor@qau.edu.pk}{mansoor@qau.edu.pk}}}
			and \textbf{Umer Zubair}$^{c}$\footnote{E-mail: \texttt{\href{mailto: umer@udel.edu}{umer@udel.edu}}}
		} 
		\vspace{.5cm}
		
		{\baselineskip 20pt \it
			$^{a}$School of Mathematics and Physics, \\ Hubei Polytechnic University, Huangshi 435003, China \\[6pt]
			$^{b}$Department of Physics, Quaid-i-Azam University,\\ Islamabad 45320, Pakistan\\[6pt]
			$^{c}$Department of Physics and Astronomy, \\ University of Delaware, Newark, DE 19716, USA\\[2pt]}
		
		\vspace{1cm}
	\end{center}

	%%%%%%%%%%%%%%%%%%%%%%%%
	\begin{abstract}
		A realistic model of $SU(5) \times U(1)_{\chi}$, embedded in $SO(10)$ supersymmetric grand unified theory, is investigated for the emergence of a metastable cosmic string network. This network eventually decays via the Schwinger production of monopole-antimonopole pairs, subsequently generating a stochastic gravitational wave background that is compatible with the NANOGrav 15-year data. In order to avoid the monopole problem in the breaking of both $SO(10)$ and $SU(5)$, a non-minimal Higgs inflation scenario is incorporated. The radiative breaking of the $U(1)_{\chi}$ symmetry at a slightly lower scale plays a pivotal role in aligning the string tension parameter with the observable range. The resultant gravitational wave spectrum not only accounts for the signal observed in the most recent pulsar timing array (PTA) experiments but is also accessible to both current and future ground-based and space-based experiments.
	\end{abstract}
	%%%%%%%%%%%%%%%%%%%%%%%%
	
	%
	\newpage
	\section{\large{\bf Introduction}}
	The recent findings presented by the NANOGrav \cite{NANOGrav:2023hvm, NANOGrav:2023gor} as well as other Pulsar Timing Array (PTA) experiments \cite{Antoniadis:2023rey, Reardon:2023gzh, Xu:2023wog, NANOGrav:2023hde} have furnished compelling evidence for the Hellings-Downs angular correlation in the shared-spectrum mechanism, previously observed through PTAs \cite{NANOGrav:2020bcs}. These results provide substantive support for the inference that the shared-spectrum phenomenon emerges from gravitational waves (GWs) within the nHz range. However, the precise origin of these GWs remains unknown, necessitating the exploration of the various scenarios and the ability to differentiate among them. While the conventional astrophysical explanation points towards GWs emitted by binary systems of supermassive black holes (SMBHs)\cite{Antoniadis:2023zhi,NANOGrav:2023hfp}, it is crucial to consider more exotic possibilities beyond the standard model of particle physics, such as cosmic strings or cosmological phase transitions \cite{NANOGrav:2023hvm}.

 Remarkably, models based on either a metastable or stable cosmic string network successfully accounted for the findings of the NANOGrav 12.5-year data  \cite{NANOGrav:2020bcs}, which had initially hinted at a common-spectrum process in the nHz range. However, the subsequent NANOGrav 15-year data \cite{NANOGrav:2023hvm, NANOGrav:2023gor} dispelled the plausibility of stable cosmic strings, leaving only the realm of metastability. This groundbreaking outcome was subsequently validated by other PTAs as well \cite{Goncharov:2021oub,Chen:2021rqp}. These findings introduce captivating avenues of exploration, shedding light on metastble cosmic strings as potential sources of stochastic gravitational waves. Consequently, they significantly contribute to the ongoing endeavor of comprehending the cosmic mechanisms responsible for their generation.

	When a symmetry group $G$ is broken down to a subgroup $H$, the resulting manifold of degenerate vacuum states, denoted as $\mathcal{M} = G/H$, plays a crucial role. The formation of defects during symmetry breaking is determined by the topology of $\mathcal{M}$, which is encoded in the homotopy groups $\pi_n(\mathcal{M})$. Specifically, the existence of topologically stable strings is associated with a nontrivial first homotopy group, $\pi_1(\mathcal{M}) \neq I$, indicating the presence of loops in $\mathcal{M}$ that cannot be contracted to a single point. Similarly, topologically stable magnetic monopoles can emerge if the second homotopy group is nontrivial, $\pi_2(\mathcal{M}) \neq I$, indicating the existence of non-contractible two-dimensional surfaces within $\mathcal{M}$.
	
	Our particular focus lies on the two-step symmetry breaking scenario, $G \rightarrow H \rightarrow K$, where the homotopy group $G/K$ is trivial, while the homotopy groups of the individual steps, $G/H$ and $H/K$, are nontrivial. This specific case allows for the formation of metastable defects. In this article, we re-examine the $SU(5)\times U(1)_\chi$ model in the light of the recent NANOGrav 15-year data. We focus on a minimal yet representative example to explore the production of metastable strings. We study the chain of symmetry breaking from $SO(10)$ to the standard model via $SU(5)\times U(1){\chi}$:
	\begin{equation*}\label{SU5B}
		SO(10) \xrightarrow{\mathbf{45}} SU(5) \times U(1)_{\chi} \xrightarrow{\mathbf{24 \oplus 16}} G_{\text{SM}} \,.
	\end{equation*}
	Specifically, we utilize $\bf{16}$-plet of $SO(10)$ for the breaking of $U(1)_\chi$. In this scenario, the homotopy group of the resulting manifold $\mathcal{M} = SO(10)/G_{\text{SM}}$ is trivial, $\pi_1(\mathcal{M}) = I$, indicating the absence of topologically stable cosmic strings. However, the formation of metastable cosmic strings becomes possible in this case \cite{Kibble:1976sj,Hindmarsh:1994re}.  Metastable cosmic strings, formed in the early universe, give rise to a stochastic gravitational-wave background (SGWB) that extends to low frequencies and encompasses a wide spectrum spanning numerous frequency decades. For recent studies on SGWB from cosmic strings, see Refs \cite{Buchmuller:2023aus, Antusch:2023zjk, Fu:2023mdu, Lazarides:2023rqf,Ellis:2020ena,King:2020hyd,Buchmuller:2020lbh,Ahmed:2021ucx,Afzal:2022vjx,Vagnozzi:2023lwo,Vagnozzi:2020gtf, Buchmuller:2021dtt, Buchmuller:2021mbb, Masoud:2021prr}. 
	
	The process of breaking the $SU(5)$ symmetry can lead to an undesired consequence: the generation of magnetic monopoles. The presence of monopoles is incompatible with observations and poses a challenge to the model. To circumvent this issue, many variants of inflation models have been proposed, such as shifted hybrid inflation \cite{Jeannerot:2002wt, Ahmed:2022wed, Khalil:2010cp}, smooth hybrid inflation \cite{Ahmed:2022vlc,Rehman:2014rpa}, Higgs inflation \cite{Arai:2011nq,Masoud:2019cen} and new inflation \cite{Rehman:2018gnr}.
	
	In the scope of our analysis, we specifically focus on the non-minimal Higgs inflation model where the $SU(5)$ gauge symmetry spontaneously breaks during inflation, inflating the monopoles away. Furthermore, the $U(1)_\chi$ symmetry is radiatively broken at some lower scale. This additional breaking mechanism plays a crucial role in adjusting the string tension parameter to align with the desired range. Aligning the string tension parameter in this manner is essential to satisfy the observational constraints set forth by the NANOGrav 15-year data and other planned experiments.

	The structure of the paper is organized as follows. In Section 2, we provide a comprehensive description of the realistic embedding of $SU(5)\times U(1)_{\chi}$ in $SO(10)$. In Section 3, we present the details of the inflationary setup and the results of the numerical analysis. Additionally, we explore the potential for observing primordial gravity waves. Moving on to Section 4, we delve into the discussion of radiative symmetry breaking and its connection to the production of a metastable cosmic string network. This network gives rise to the generation of a stochastic gravitational wave background (SGWB), which has been confirmed by NANOGrav 15-year data. Finally, in Section 5, we summarize our findings and draw our conclusions.
	%
	%%%%%%%%%%%%%%%%%%%%%%%%%%%%%%%%%%%%%%%%%%%%%%%%%%%%%%%
	\section{The $SU(5) \times U(1)_{\chi} \subset SO(10)$ Embedding} \label{sec2}%
	
	The subgroup $SU(5) \times U(1)_{\chi}$ is embedded within the larger group $SO(10) \supset SU(5) \times U(1)_{\chi}$, where the decomposition of the various multiplets of $SO(10)$ can be written as
	\begin{eqnarray}
		\mathbf{10} &=& (\mathbf{5}, 2) + (\bar{\mathbf{5}}, -2), \\
		\mathbf{16} &=& (\mathbf{10}, -1) + (\bar{\mathbf{5}}, 3) + (\mathbf{1}, -5),\\
		\mathbf{45} &=& (\mathbf{24}, 0) + (\mathbf{10}, 4) + (\bar{\mathbf{10}}, -4) + (\mathbf{1}, 0).
	\end{eqnarray}
	The MSSM matter content and the right-handed neutrino (RHN) superfield reside in the $\mathbf{16}$ (spinorial) representation of $SO(10)$ which is further decomposed into $(\mathbf{10}, -1)$, $(\bar{\mathbf{5}}, 3)$ and $(\mathbf{1}, -5)$ representations of $SU(5) \times U(1)_{\chi}$. The fundamental representation, $\mathbf{10}_{1H}$, of $SO(10)$ contains both the electroweak Higgs doublets, ($h_{u}, h_{d}$), and the color Higgs triplets, ($D_{h},\bar{D}_{\bar{h}}$), Given that the adjoint representation $\mathbf{45}$ belongs to a rank-two antisymmetric tensor, it requires the inclusion of an additional fundamental Higgs ($\mathbf{10}_{2H}$) in order to facilitate a non-zero interaction between $\mathbf{45}_H$ and two 10-plets. This interaction is an important ingredient in the later-discussed doublet-triplet splitting. The breaking of $SO(10)$ symmetry down to SM symmetry is realized by the non-zero vacuum expectation values (VEVs) of $\mathbf{45}_H$ and $\mathbf{16}_H$, $\widebar{\mathbf{16}}_H$:
	\begin{equation} \label{eq:breaking_pattern}
		SO(10) \xrightarrow{\langle \mathbf{45} \rangle} SU(5) \times U(1)_{\chi} \xrightarrow{\langle (\mathbf{24}, 0) \rangle} G_{\text{SM}} \times U(1)_{\chi} \xrightarrow{\langle \mathbf{16}, \widebar{\mathbf{16}} \rangle} G_{\text{SM}}.
	\end{equation}
	Here, the $SU(5)$ symmetry is broken by VEV of ($\Phi \equiv (\mathbf{24}, 0)$) in $\mathbf{45}_H$, while $U(1){\chi}$ is broken by the VEV of ($\widebar{\mathbf{16}}_H, \mathbf{16}_H$) in the RHN direction ($\nu_H, \bar{\nu}_H$).
	The breaking of $SO(10)$ and $SU(5)$ symmetry yields stable magnetic monopoles, while the breaking of $U(1)_{\chi}$ yields metastable cosmic strings \cite{Kibble:1982ae}.  
	\begin{table}[!htb]
		\setlength\extrarowheight{3pt}
		\centering
		\begin{tabular}{c c c c}
			\hline \hline \rowcolor{Gray}
			\multicolumn{1}{c}{}                              & \multicolumn{1}{c}{}                                                                                        & \multicolumn{1}{c}{}                                                                        \\ \rowcolor{Gray}
			\multicolumn{1}{c}{\multirow{-2}{*}{\begin{tabular}[c]{@{}c@{}}$SO(10)$\\ \textbf{Representations}\end{tabular}}} & 
			\multicolumn{1}{c}{\multirow{-2}{*}{\begin{tabular}[c]{@{}c@{}}\textbf{Global}\\ $U(1)_R $\end{tabular}}} &
			\multicolumn{1}{c}{\multirow{-2}{*}{\begin{tabular}[c]{@{}c@{}}\textbf{Decomposition under}\\ $SU(5) \times U(1)_{\chi}$\end{tabular}}} &  \multicolumn{1}{c}{\multirow{-2}{*}{\begin{tabular}[c]{@{}c@{}}\textbf{Decomposition under}\\ $G_{\text{SM}}$\end{tabular}}}  \\  \hline
			\rowcolor{Gray2} 	\multicolumn{4}{c}{\textbf{Matter sector}}                                      \\ \hline
			\multicolumn{1}{c}{$\mathbf{16}_i$}     &  1/2 &    $F_i \left( \mathbf{10}, -1 \right)$  &       $Q_i \left( \mathbf{3}, \mathbf{2}, 1/6 \right)$   \\
			\multicolumn{1}{c}{}     &     &                                          &       $u^c_i\left( \bar{\mathbf{3}}, \mathbf{1}, -2/3 \right)$     \\
			\multicolumn{1}{c}{}     &     &                                          &       $e^c_i\left( \mathbf{1}, \mathbf{1}, 1\right)$     \\[5pt]
			\multicolumn{1}{c}{} &    &        $\bar{f}_i \left( \bar{\mathbf{5}}, 3 \right)$ & $d^c_i\left( \bar{\mathbf{3}}, \mathbf{1}, 1/3 \right)$  \\
			\multicolumn{1}{c}{} &      &                                             & $\ell_i\left( \mathbf{1}, \mathbf{2}, -1/2\right)$   \\[5pt]
			\multicolumn{1}{c}{}	&        &    $\nu_{i}^c \left( \mathbf{1}, -5 \right)$ & $\nu^c_i\left( \mathbf{1}, \mathbf{1}, 0 \right)$  \\[5pt] \hline 
			\rowcolor{Gray2} 	\multicolumn{4}{c}{\textbf{Higgs sector}}                               \\ \hline
			\multicolumn{1}{c}{$\mathbf{45}_H$}	&   0   &    $\Phi \left( \mathbf{24}, 0 \right)$ & $\Phi_{24}(\mathbf{1}, \mathbf{1}, 0)$ \\
			\multicolumn{1}{c}{ }    	&         &                          & $W_H (\mathbf{1}, \mathbf{3}, 0)$         \\
			\multicolumn{1}{c}{ }    	&         &                          & $G_H (\mathbf{8}, \mathbf{1}, 0)$         \\
			\multicolumn{1}{c}{ }    	&         &                          & $Q_H(\mathbf{3}, \mathbf{2}, -5/6)$         \\
			\multicolumn{1}{c}{ }    	&         &                          & $\bar{Q}_H(\mathbf{3}, \mathbf{2}, 5/6)$         \\[5pt]
			\multicolumn{1}{c}{$\mathbf{10}_{1H}$}	&     0    &   $h_1\left( \mathbf{5}, 2 \right)$ & $D_h(\mathbf{3}, \mathbf{1}, -1/3)$  \\
			\multicolumn{1}{c}{   }	&           &                       & $h_u (\mathbf{1}, \mathbf{2}, 1/2)$       \\[5pt]
			\multicolumn{1}{c}{} &      &$\bar{h}_1\left( \bar{\mathbf{5}}, -2 \right)$ & $\bar{D}_{\bar{h}}(\bar{\mathbf{3}}, \mathbf{1}, 1/3)$ \\
			\multicolumn{1}{c}{         } &       &                                & $h_d(\mathbf{1}, \mathbf{2}, -1/2)$                \\[5pt]
			\multicolumn{1}{c}{$\mathbf{10}_{2H}$}	&  1  &  $h_2\left( \mathbf{5}, 2 \right)$ & $D^{'}_h(\mathbf{3}, \mathbf{1}, -1/3)$  \\
			\multicolumn{1}{c}{   }	&        &                          & $h^{'}_u (\mathbf{1}, \mathbf{2}, 1/2)$      \\[5pt]
			\multicolumn{1}{c}{} &           &$\bar{h}_2\left( \bar{\mathbf{5}}, -2 \right)$ & $\bar{D}^{'}_{\bar{h}}(\bar{\mathbf{3}}, \mathbf{1}, 1/3)$ \\
			\multicolumn{1}{c}{         } &        &                               & $h^{'}_d(\mathbf{1}, \mathbf{2}, -1/2)$                          \\[5pt]
			\multicolumn{1}{c}{$\mathbf{16}_H$}	&   0    &  $\nu_H\left( \mathbf{1}, 5 \right)$ & $\nu_H \left( \mathbf{1}, \mathbf{1}, 0\right)$ \\
			\multicolumn{1}{c}{$\widebar{\mathbf{16}}_H$} &      0     & $\bar{\nu}_H\left( \mathbf{1}, -5 \right)$ & $\bar{\nu}_H \left( \mathbf{1}, \mathbf{1}, 0\right)$ \\
			\multicolumn{1}{c}{$\mathbf{1}$}	&      1      &  $S \left( \mathbf{1}, 0 \right)$ & $S \left( \mathbf{1}, \mathbf{1}, 0\right)$ \\[3pt]  \hline \hline
		\end{tabular}
		\caption{The decomposition of matter and Higgs representations of $SO(10)$ under $SU(5) \times U(1)_{\chi}$ and $G_{\text{SM}}$ along with their global $U(1)_R$ charges.}
		\label{tab:field_charges}
	\end{table}
	
	The decomposition of $SO(10)$ representations employed in this model under the $SU(5) \times U(1)_{\chi}$ and SM gauge group is provided in Table \ref{tab:field_charges}, along with their $U(1)_R$ charge assignments. Assuming $U(1)_R$ symmetry at the renormalizable level, the superpotential of the $SO(10)$ model can be written as
	\begin{equation}
		W_{SO(10)} = W_\text{DT-Splitting} + W_\text{Yukawa} + W_{\text{Inflation}} + W_{SO(10)\text{-Breaking}},
	\end{equation}
	with,
	\begin{eqnarray}
		W_\text{DT-Splitting} &=& \gamma \mathbf{10}_{1H} \mathbf{45}_H \mathbf{10}_{2H} + M_{12} \mathbf{10}_{2H} \mathbf{10}_{1H} + \frac{M_{22}}{\Lambda^2} \mathbf{45}_H^2 \mathbf{10}_{2H} \mathbf{10}_{2H}, \label{W_DTS} \\
		W_\text{Yukawa} &=& y^f_{ij} \mathbf{16}_i \mathbf{10}_{1H} \mathbf{16}_j + \frac{\lambda_{ij}}{\Lambda} \mathbf{16}_i \mathbf{16}_j \bar{\mathbf{16}}_H\bar{\mathbf{16}}_H, \label{W_yukawa}\\
		W_{\text{Inflation}} &=&  S \left(\kappa M_{24} - \kappa \mathbf{45}_H^2 +  \sigma_{16} \bar{\mathbf{16}}_H \mathbf{16}_H + \lambda \mathbf{10}_{1H} \mathbf{10}_{1H}  \right), \label{W_inf}\\
		W_{SO(10)\text{-Breaking}} &=& \frac{1}{\Lambda^2} \left(  M_{45} \mathbf{45}_H^4 + \lambda_{45} \mathbf{45}_H^5 \right) \label{W_so10_breaking}.
	\end{eqnarray}
Given that $U(1)_R$ is a global symmetry, its potential violation at the nonrenormalizable level is anticipated \cite{Nelson:1993nf, Civiletti:2013cra}. Exploiting this understanding, we incorporate $W_{SO(10)\text{-Breaking}}$ which is relevant for $SO(10)$ breaking into $SU(5) \times U(1)_{\chi}$. This breaking is realized via the nonzero VEV of $\mathbf{45}_H$ in the following direction
	\begin{equation}
		\langle \mathbf{45}_H \rangle = \begin{pmatrix}
			0 & 1\\
			-1 & 0 
		\end{pmatrix} \otimes \begin{pmatrix}
			a & 0 & 0 & 0 & 0\\
			0 & a & 0 & 0 & 0\\
			0 & 0 & a & 0 & 0\\
			0 & 0 & 0 & a & 0\\
			0 & 0 & 0 & 0 & a
		\end{pmatrix} ,
	\end{equation} 
	with $\langle \mathbf{45}_H^2 \rangle = -10\, a^2$, $\langle \mathbf{45}_H^4 \rangle = 10\, a^4, \, 100 a^4$ and $\langle \mathbf{45}_H^5 \rangle = 3840\, a^5$. Taking $a \sim M_{45} / \lambda_{45}$, the $SO(10)$ symmetry breaking scale can be adjusted above the $SU(5) \times U(1)_{\chi}$ breaking scale $M_{\text{GUT}} \sim 2 \times  10^{16}$ GeV and below $M_{\text{string}} \sim 5 \times 10^{17}$~GeV.

	The superpotential $W_\text{DT-Splitting}$ contains terms relevant for the solution of doublet-triplet splitting problem. After the $SO(10)$ symmetry breaking, the first two terms in \eqref{W_DTS} yield;
	\begin{align}
		\gamma \mathbf{10}_{1H} \mathbf{45}_H \mathbf{10}_{2H} + M_{12} \mathbf{10}_{2H} \mathbf{10}_{1H} &\supset \gamma \left[\left( \bar{h}_2 \Phi h_1 + h_2 \Phi \bar{h}_1 \right) +  \langle (\mathbf{1}, 0)_{\mathbf{45}_H} \rangle \left( \bar{h}_2 h_1 + h_2 \bar{h}_1 \right)\right] \nonumber \\
		&+ M_{12} \left( \bar{h}_2 \Phi h_1 + h_2 \Phi \bar{h}_1 \right),
	\end{align}
	which generates heavy masses for the color-triplets ($D_{h},\bar{D}_{\bar{h}}, D^{'}_{h},\bar{D}^{'}_{\bar{h}}$) after $SU(5)$ symmetry breaking and light masses for the respective electroweak doublets by fine-tuning of the parameters involved. To render the extra pair of electroweak doublets, ($h^{'}_{u}, h^{'}_{d}$), heavy, the following $R$ symmetry breaking term at nonrenormalizable level is introduced in \eqref{W_DTS}:
	\begin{equation}
		\frac{M_{22}}{\Lambda^2} \mathbf{45}_H^2 \mathbf{10}_{2H} \mathbf{10}_{2H} \supset \frac{M_{22}}{\Lambda^2} \Tr{(\Phi^2)} \bar{h}_2 h_2.
	\end{equation}
As a result of this addition, the electroweak doublets ($h^{'}_{u}, h^{'}_{d}$) contained within $h_2$, $\bar{h}_2$, acquire heavy masses after $SU(5)$ symmetry breaking. The last term in \eqref{W_inf} involving the mixing of electroweak doublets in ($h_1$, $\bar{h}_1$) gives rise to the $\mu$-term in MSSM, where $\mu \sim \lambda \langle S \rangle $. As shown in \cite{Dvali:1997uq}, the soft susy breaking terms can yield $\langle S \rangle \sim m_{3/2}/ \kappa$ and the $\mu$-problem is solved with $\mu \sim (\lambda / \kappa) m_{3/2}$.
	
	The superpotential for the $SU(5)\times U(1)_{\chi}$ model is achieved by retaining the relevant terms from  $W_{SO(10)}$,   
	\begin{eqnarray}
		W_{SU(5)\times U(1)_{\chi}} &=& \kappa S \left( M^2 - \Tr (\Phi^2) \right) + \lambda S h_1 \bar{h}_1 + \sigma_{\chi} S \nu_H \bar{\nu}_H \nonumber \\
		&+& y_{ij}^{(u)} F_i F_j h_1 + y_{ij}^{(d,e)} F_i \bar{f}_j \bar{h}_1 	+ y_{ij}^{(\nu)} \nu_{i}^c \bar{f}_j h_1 +\frac{\lambda_{ij} }{\Lambda} \nu_H \nu_H \nu_{i}^{c} \nu_{j}^{c} \nonumber \\
		&+& \gamma_{12} \left( \bar{h}_2 \Phi h_1 + h_2 \Phi \bar{h}_1 \right) + \delta_{12} \left( \bar{h}_2 h_1 + h_2 \bar{h}_1 \right) + \delta_{22} \bar{h}_2 h_2 ,
	\end{eqnarray}
	where $M$ is a superheavy mass of the order of GUT scale and $\Lambda$ is the cutoff scale between the GUT scale ($2 \times 10^{16}$ GeV) and reduced Planck scale $m_P$. The first two terms in the first line are relevant for Higgs inflation  while, the Yukawa couplings $y_{ij}^{(u)}$, $ y_{ij}^{(d,e)}$, $y_{ij}^{(\nu)}$ of order $y^f$ in the second line generate Dirac masses for quarks and leptons after the electroweak symmetry breaking. For a realistic construction of the mass spectrum of charged fermions within $SO(10)$, refer to \cite{Albright:1997xw, Albright:1998sy, Kyae:2005vg}. The first two terms in the third line are involved in the solution of doublet-triplet splitting, where the four color-triplets in ($h_1, \bar{h}_1, h_2, \bar{h}_2$) mix to acquire masses of order $\sim M$. The last term arises at non-renormalizable level with $R$-symmetry breaking, where the mixing between doublet pair in ($h_2, \bar{h}_2$) yields their masses of order $\sim M^2/\Lambda$. Furthermore, the right-handed neutrino mass matrix, $m_{\nu_{ij}} = \frac{\lambda_{ij} }{\Lambda} \langle \nu_H \rangle^2$, is generated after $\nu_H$ acquires a VEV through radiative breaking of $U(1)_{\chi}$ symmetry. This mass matrix along with the Dirac mass matrix arising from $y_{ij}^{(\nu)}$ Yukawa coupling can generate tiny neutrino masses.
 
As a consequence of $R$ symmetry, some components of the Higgs multiplets remain light \cite{Barr:2005xya, Fallbacher:2011xg}. For instance, some components in the multiplets of $\mathbf{45}_H$, $\mathbf{16}_H$ and $\widebar{\mathbf{16}}_H$ representations remain light and may spoil gauge coupling unification. These fields can however acquire  masses order of $10^{14}$~GeV through the introduction of $U(1)_R$ breaking non-renormalizable terms. Thus, the restoration of gauge coupling unification can be achieved through the introduction of a mass splitting within a few appropriate supplementary multiplets at the GUT scale \cite{Khalil:2010cp, Masoud:2019gxx, Ahmed:2022thr}.
	
	\section{Non-Minimal Higgs Inflation}\label{model1}
	The superpotential terms relevant for Higgs inflation are
	\begin{equation}
		W \supset  \kappa S \left(  M^2 -  \frac{1}{2} \sum_{i}\phi_{i}^{2}   \right)  .
	\end{equation}
	In non-minimal Higgs inflation\footnote{For implementation of Higgs inflation in other supersymmetric GUT models, see refs \cite{Masoud:2019cen, Ahmed:2021dvo, Ahmed:2018jlv}}, the adjoint Higgs field $\Phi$ plays the role of inflation. To realize non-minimal Higgs inflation in $SU(5) \times U(1)_{\chi}$ with $R$-symmetry, we assume the following no-scale like form of the K\"{a}hler potential,
	\begin{eqnarray} \label{K}
		K = -3 \log  \left( 1 -\frac{1}{3}\left( \lvert S \rvert ^2 + \Tr{(\lvert \Phi \rvert ^2)}  \right)  +  \frac{1}{2} \delta( \Tr{(\Phi^2)} + \text{h.c.})+\frac{1}{3}\gamma\lvert S\lvert^4 + \cdots\right),
	\end{eqnarray}
	where we have assumed the stabilization of the modulus fields \cite{  Ahmed:2022wed, Ahmed:2018jlv,Cicoli:2013rwa, Ellis:2013nxa} and adopt units where the reduced Planck mass, $m_P = 1$. In the exact no-scale limit, the couplings $\delta$ and $\gamma$ vanish. The term with the coupling $\delta$ plays a pivotal role in non-minimal Higgs inflation, while the higher-order term with the coupling $\gamma$ is required to stabilize $S$ during inflation \cite{Lee:2010hj}. The VEV’s of the fields at the global SUSY minimum are given by
	\begin{gather}     
		\langle \phi_i  \rangle = \sqrt{2} M, \qquad \langle S \rangle = 0, 
	\end{gather}
	where the gauge symmetry breaking scale $M$ is taken to be  the order of GUT scale, $M_{G}\equiv 2\times 10^{16}$ GeV. The scalar potential in the Einstein' frame is defined as, 
	\begin{equation}
		V_E = e^{G}\left( G_i (G^{-1})_j^i  G^j- 3\right)+\frac{1}{2} g_a^2 G^i (T_a)_i^j z_j (\text{Re} f^{-1}_{ab}) G^k (T_b)_k^l z_l,
	\end{equation}
	where
	\begin{equation}
		G = K + \log|W|^2, \quad  G^i \equiv \frac{\partial G}{\partial z_i} ,\quad G_i = \frac{\partial G}{\partial z_i^*}, \quad G^i_j = \frac{\partial G}{\partial z_i \partial z_j^*},
	\end{equation}
	with $z_i \in {S, \Phi}$, $f_{ab}$ is the gauge kinetic function and $T_a$ are the generators of the gauge group. After the stabilization of $S$ in their appropriate vacua, the scalar potential in Einstein' frame can be written as, 
	\begin{eqnarray}
		V_E = \dfrac{\kappa^2 \left| \frac{1}{2} \phi_{24}^{2} - M^2  \right|^2 }{\left( 1-\frac{1}{6} \lvert \phi_{24} \rvert^2  + \frac{\delta}{4}\left( \phi_{24}^2 + \text{h.c.}\right) \right)^2}, 	%+ \frac{g_5^2}{2}   \frac{ \left( f^{ijk} \phi_j \phi_{k}^{\dagger}  \right)^2}{\left( 1-\frac{1}{6} \lvert \phi_{i} \rvert^2  + \frac{\delta}{4}\left( \phi_{i}^2 + \text{h.c.}\right) \right)^2},
	\end{eqnarray}
	where we have rotated the adjoint Higgs field $\Phi = \phi_i T^i$ in the $24$-direction,
	\begin{equation}
		\Phi  = \frac{\phi_{24}}{\sqrt{15}} \left( 1, 1, 1, - 3/2, - 3/2 \right),
	\end{equation}
	using $SU(5)$ gauge transformation, which also leads to the $D$-flatness condition. Note that, the fields $\nu_H$, $\bar{\nu}_H$ acquire heavy masses from the superpotential terms $(\mathbf{16}_H \widebar{\mathbf{16}}_H)^2 + \mathbf{16}_H \mathbf{45}_H^2 \widebar{\mathbf{16}}_H$, and quickly stabilize at $\nu_H = \bar{\nu}_H = 0$. Rotating the complex scalar field $\phi_{24}$ to the real axis $\phi_{24} = h/\sqrt{2}$, the scalar potential in Einstein’ frame along the $D$-flat direction becomes
	\begin{eqnarray}
		V_E = \frac{\kappa^2 \left( h^{2} - 4 M^2  \right)^2 }{16 \left( 1 + \xi h^2 \right)^2} = \frac{\kappa^2 \left( h^{2} - 4 M^2  \right)^2 }{16\, \Omega^2}, \quad \text{with} \quad \xi = \frac{\delta}{4} - \frac{1}{12}, 
	\end{eqnarray}
	where, the conformal scaling factor, $\Omega$, which relates the Einstein and Jordan frames, is given by
	\begin{equation}
		g^J_{\mu\nu} = \Omega g^E_{\mu\nu} = (1+\xi h^2) g^E_{\mu\nu}.
	\end{equation}
	The Lagrangian in the Einstein' frame is
	\begin{equation}\label{Jordan}
		\mathcal{L}_E=\sqrt{-g_E}\left[\dfrac{1}{2}\mathcal{R}(g_E) - \frac{1}{2} g_E^{\mu \nu} \partial_\mu \hat{h} \partial_\nu \hat{h} - V_E (\hat{h}(h))\right],
	\end{equation}
	where $\hat{h}$ is the canonically normalized inflaton field defined as,
	\begin{equation}
		\frac{d\hat{h}}{dh} \equiv J = \sqrt{\dfrac{\Omega+6\xi^2h^2}{\Omega^2}}.
	\end{equation}
	
	The slow-roll parameters can be expressed in terms of $h$ as
	\begin{eqnarray}
		\epsilon(h)&=&\frac{1}{2J^2}\left(\dfrac{V_E^\prime (h)}{V_E }\right)^2, \quad
		\eta(h) = \frac{1}{J^2} \left( \frac{V_E^{\prime \prime}(h)}{V_E} -  \sqrt{2 \epsilon} J^{\prime}(h) \right),  \label{epsilon} \\
		\zeta^2(h) &=& \frac{\sqrt{2 \epsilon}}{J^3} \left(\dfrac{V_E^{\prime \prime \prime}(h)}{V_E} - 3 \eta J J^{\prime}(h) - \sqrt{2 \epsilon} J^{\prime \prime}(h)\right), 
	\end{eqnarray}
	where primes denote the derivatives with respect to $h$. 
	The scalar spectral index $n_s$, the tensor to scalar ratio $r$ and the running of the scalar spectral index $\frac{d n_s}{d\ln k}$ are given, to leading order in slow-roll parameters, as
	\begin{eqnarray}
		n_s &\simeq& 1-6 \epsilon(h_0) + 2 \eta(h_0), \label{ns}\\
		r &\simeq& 16 \epsilon(h_0), \label{r} \\
		\frac{d n_s}{d\ln k} &\simeq& 16\epsilon(h_0)\eta(h_0) -24 \epsilon^2(h_0)-2\zeta^2(h_0),  \label{alfa}
	\end{eqnarray}
	where $h_0$ is the field value at the pivot scale. The amplitude of the scalar power spectrum is given by 
	\begin{equation}\label{As}
		A_s(k_0)=\left. \frac{1}{{24} \pi^2 }\frac{V_E(h)}{\epsilon(h)}\right|_{h(k_0) = h_0},
	\end{equation}
	The last $N_0$ number of e-folds from $h=h_0$ to the end of inflation at $h=h_e$ is expressed as 
	\begin{equation} \label{N0}
		N_0=\int_{h_e}^{h_0}  \dfrac{J(h)}{\sqrt{2\epsilon(h)}} dh.
	\end{equation}
	Assuming a standard thermal history, $N_0$ is related to $T_r$ as \cite{Kolb:1990vq},
	\begin{equation} \label{Nth}
		N_{0}\simeq 53 + \frac{1}{3} \ln\left(\frac{T_{r}}{10^{9}\text{ GeV}}\right)+\frac{2}{3}\ln\left(  \frac{\sqrt{\kappa} M}{10^{15} \text{ GeV}}\right).
	\end{equation}
	The reheat temperature $T_{r}$ can be estimated as
	\begin{equation}\label{reheat}
		T_{r}\simeq\sqrt[4]{\frac{72}{5\pi^2{g_*}}} \sqrt{\Gamma_S+\Gamma_{y_{t}} }\ ,
	\end{equation}
	where $g_*=228.75$ for MSSM. The $\mu$-term $\lambda S h_1 \bar{h}_1$\footnote{A $\mu$-term with a value around a few hundred GeV requires a gravitino mass in the range $(4.4 \times 10^5 - 1.8 \times 10^8)$ GeV, for the corresponding ranges of $\kappa$ and $\lambda$.} induces an inflaton decay into Higgsinos with a decay width given by \cite{Ahmed:2022wed, Pallis:2018ver, Pallis:2018acu},
	\begin{equation}\label{inf_decay_width}
		\Gamma_S(S \rightarrow \widetilde h_u\widetilde h_d)=\frac{\lambda ^2 }{8 \pi \Omega_0^2}m_{\text{inf}},
	\end{equation}
	where $m_{\text{inf}} =  \sqrt{2} \kappa M/\Omega_0 J_0$ is the inflaton mass with
	\begin{equation}
		J_0 = \sqrt{\frac{1}{\Omega_0} + \frac{6 \xi^2 M^2}{\Omega_0^2}}, \qquad \Omega_0= 1 + 4 \xi M^2.
	\end{equation}
	 Additionally, there are two other channels that allow inflaton decay i.e  $\sigma_{\chi} S \nu_H \bar{\nu}_H$ and via the top Yukawa coupling, $y_{3 3}^{(u,\nu)} Q_3 L_3 H_u$, in a supergravity framework as described in \cite{Endo:2006qk,Endo:2007sz}. In no-scale like SUGRA models, this decay width is given by \cite{Pallis:2011gr},
\begin{equation}
\Gamma_{y_{t}}=\frac{3}{128 \pi^3} \left( \frac{6 \, \xi \, y_{t}  \, \Omega_0^{3/2}}{J_0} \right)^2  
\left( \frac{M}{m_P} \right)^2 \left( \frac{m_\text{inf}}{m_P} \right)^2 m_\text{inf},
\end{equation}
where, $y_t=y_{3 3}^{(u,\nu)}$ corresponds to the top Yukawa coupling. Note that the inflaton decay into $\nu_H$, $\bar{\nu}_H$ is kinematically forbidden as the mass of $\nu_H$ field is greater than that of the inflaton. However, the contribution of the Higgino and Yukawa coupling channels to the parametric space depends on the specific values of the Yukawa coupling and the coupling $\lambda$. To avoid the gravitino problem \cite{Ellis:1984eq, Khlopov:1984pf}, the reheating temperature is fixed at $\sim 10^9$ GeV in the case of the Higgino channel, which requires $\lambda \sim 10^{-6}$ for $77\leq\xi\leq 236$. As $\xi$ increases further, the Yukawa channel dominates over the Higgsino channel. 

\begin{figure}[t]
		\centering \includegraphics[width=8.035cm]{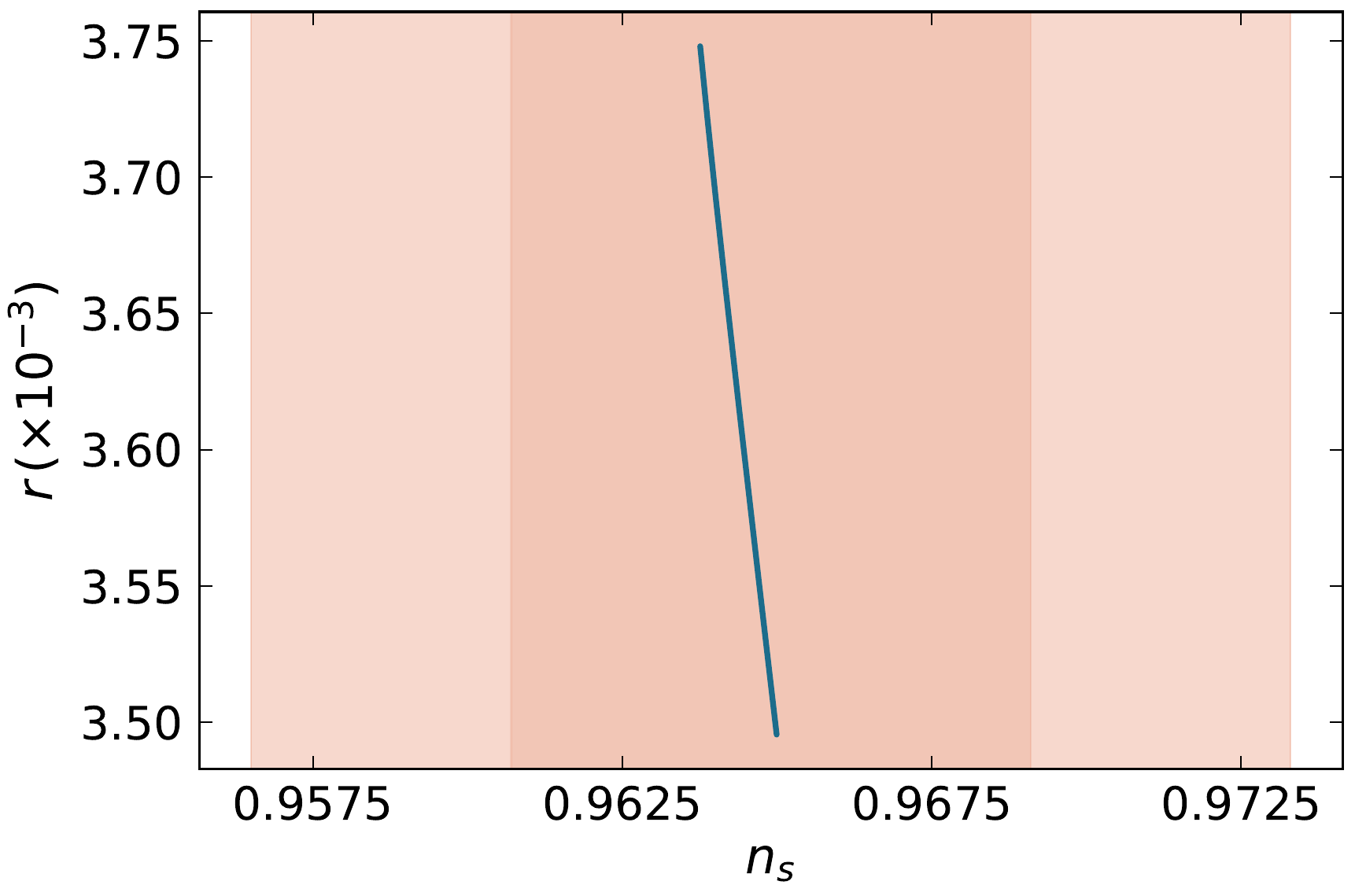}
		\centering \includegraphics[width=8.035cm]{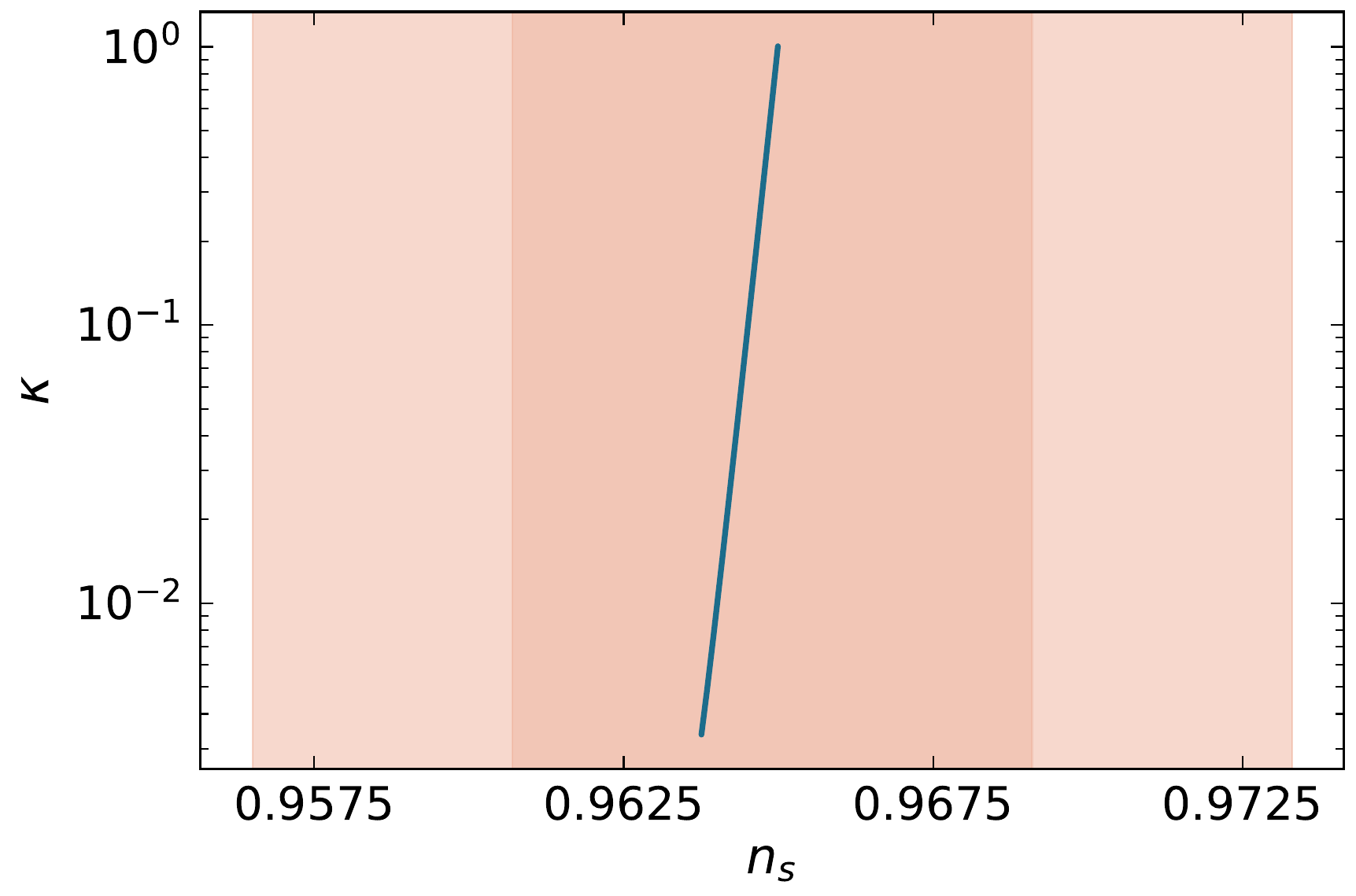}
		\caption{The variation of tensor-to-scalar ratio $r$ (left) and $\kappa$ (right) with the scalar spectral index $n_s$. The dark (light) shaded region represents the Planck 2018 1-$\sigma$ (2-$\sigma$) bounds \cite{Planck:2018jri}.}
		\label{fig:ns_r_k}
	\end{figure}

\begin{figure}[!htb]
		\centering \includegraphics[width=8.035cm]{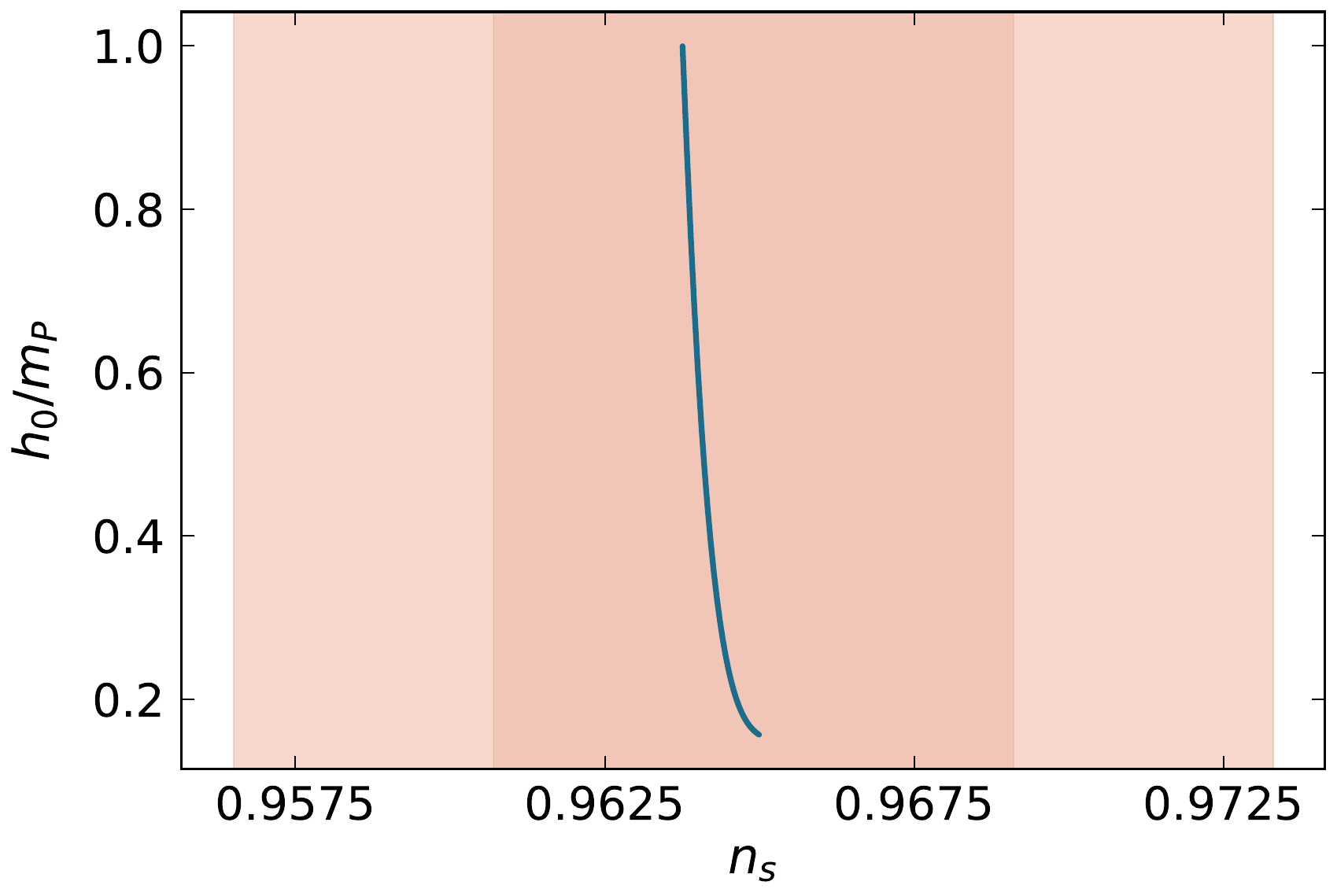}
		\centering \includegraphics[width=8.035cm]{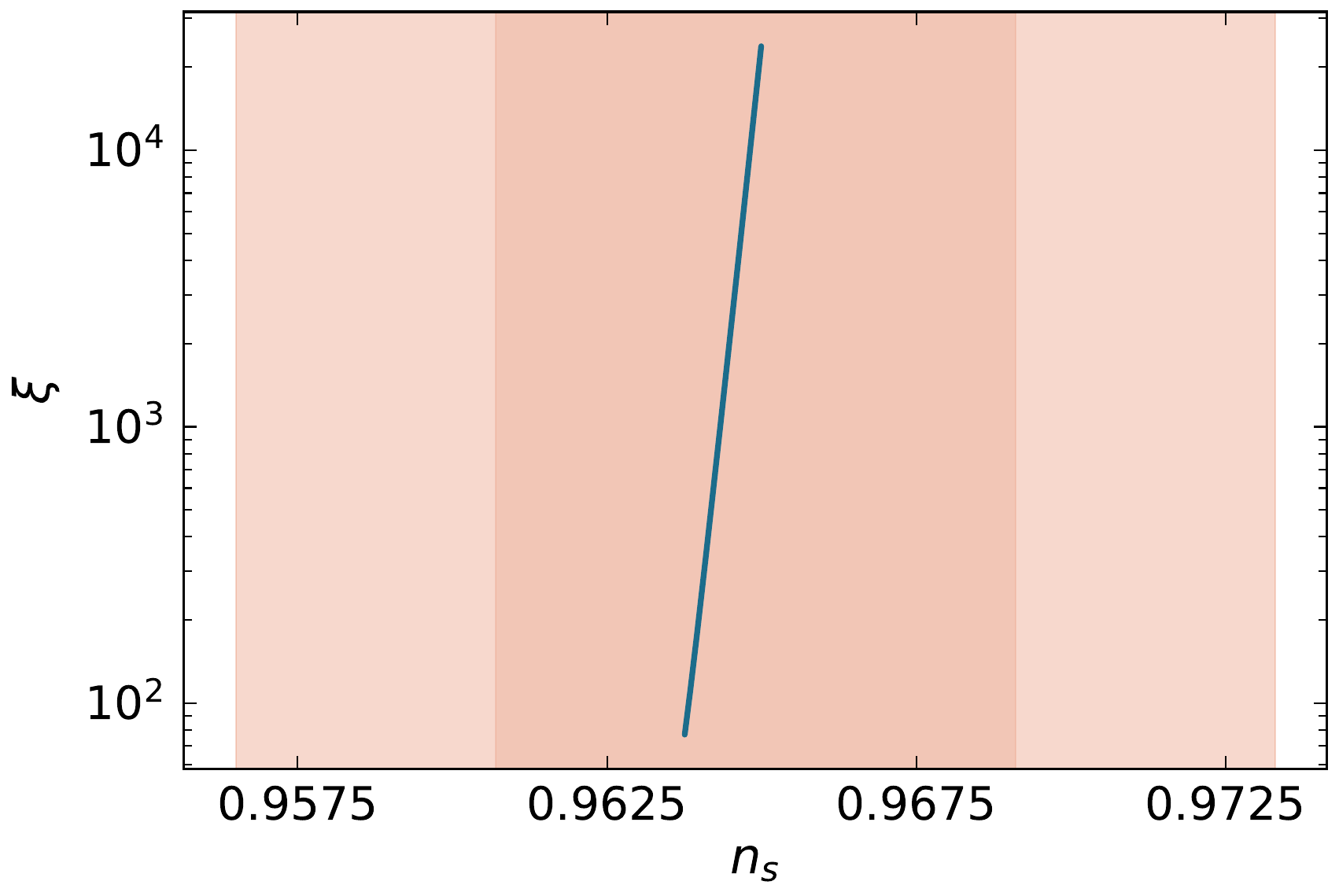}
		\caption{The variation of the inflaton field value $h_0/m_P$ (left) and  the non-minimal coupling $\xi$ (right) with the scalar spectral index $n_s$. The dark (light) shaded region represents the Planck 2018 1-$\sigma$ (2-$\sigma$) bounds.}
		\label{fig:ns_h0_xi}
	\end{figure}

\begin{figure}[!htb]
		\centering \includegraphics[width=8.035cm]{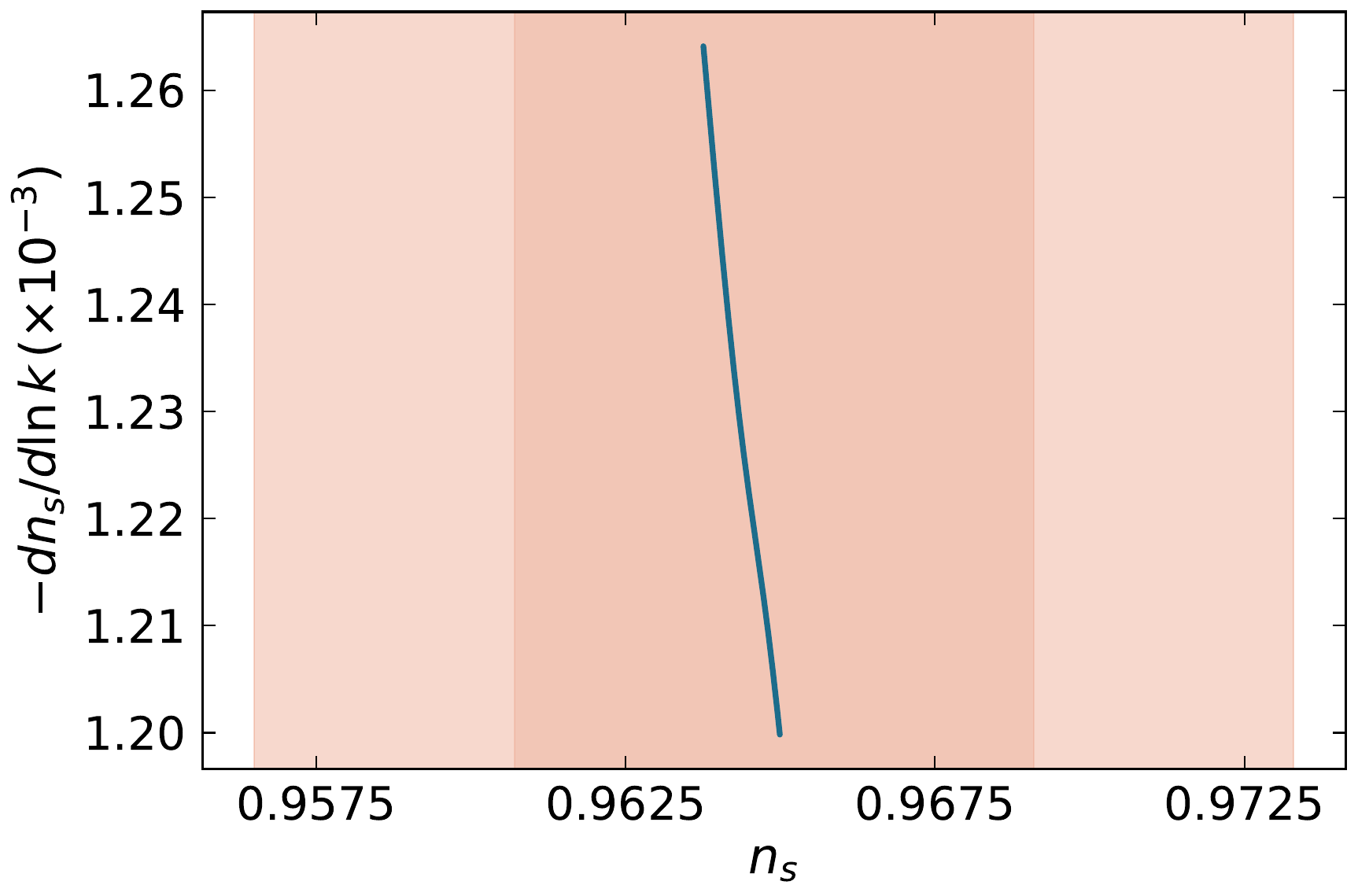}
		\centering \includegraphics[width=8.035cm]{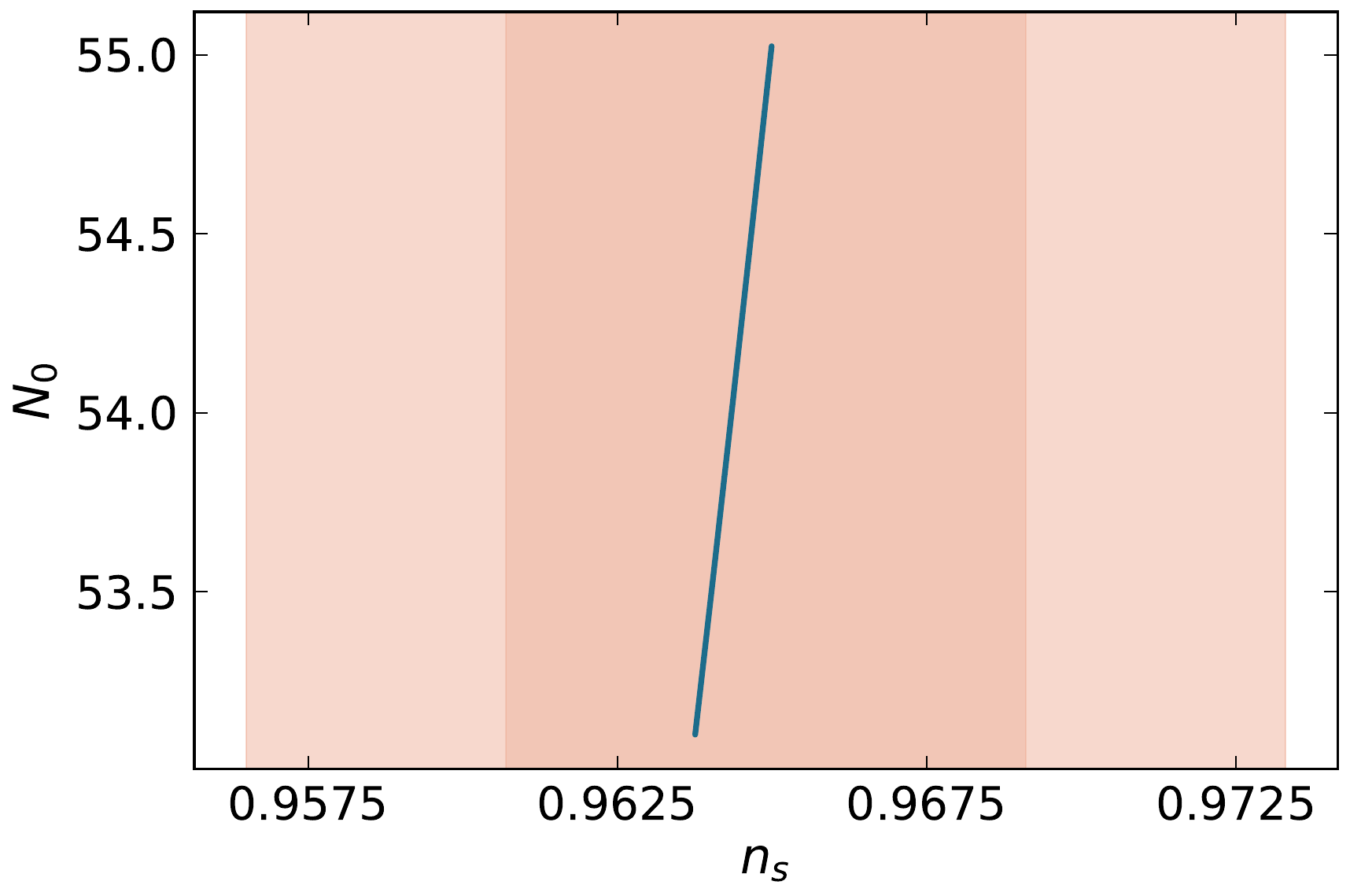}
		\caption{The variation of running of the scalar spectral index $dn_s/d \ln k$ (left) and  the number of e-folds $N_0$ (right) with the scalar spectral index $n_s$. The dark (light) shaded region represents the Planck 2018 1-$\sigma$ (2-$\sigma$) bounds.}
		\label{fig:ns_dns_n0}
	\end{figure}

The results of our numerical calculations are displayed in figures \ref{fig:ns_r_k}-\ref{fig:ns_dns_n0} where the behavior of various parameters is plotted against the scalar spectral index $n_s$ within Planck 2018 2-$\sigma$ bounds. The radiative corrections along with the soft SUSY breaking terms are negligible in non-minimal Higgs inflation.
	
	Using the Planck normalization constraint on $A_s$, $\kappa$ can be expressed in terms of $h_0$ as,
	\begin{equation} \label{kxi1}
		\kappa \simeq \frac{16 \pi \sqrt{2 A_s(k_0)}  ( 1 + 4 \xi M^2 ) (1 + \xi h_0^2)}{|\xi|(h_0^2 - 4 M^2)^2}.
	\end{equation} 
	Using Eq. \eqref{N0}, the field value $h_0$ can be expressed in terms of $N_0$ as
	\begin{equation}  \label{h0}
		h_0 \simeq \sqrt{h_e^2 + \frac{4 N_0 (1 + 4 \xi M^2 )}{3 \xi}},
	\end{equation}
	where the field value $h_e$ is obtained from the condition $\epsilon(h_e)=1$,
	\begin{equation}
		h_e \simeq \sqrt{4 M^2 + \frac{ 2(1 + 4 \xi M^2)}{\sqrt{3}\,\xi}}.
	\end{equation}
	In the leading order slow-roll approximation and in the large $\xi$ limit with $|\xi|M^2 \ll 1$, we obtain the following analytic expressions for the scalar spectral index $n_s$ and tensor-to-scalar-ratio $r$,
	\begin{align}\label{ns_analytic}
		n_s &\simeq 1 - 2 \dfrac{(1 + 4 \xi M^2)\left[ 4 M^2 - 3(1 + 4 \xi M^2)h_0^2 -12(1 + 8 \xi M^2)\xi^2 h_0^4 + 12 \xi^3 h_0^6 \right]}{9\,\xi^4 h_0^4\left(h_0^2 - 4 M^2\right)^2}, \\ \label{r_analytic}
		r &\simeq \frac{64 \left( 1 + 8 \xi M^2  \right)}{3 \left(h_0^2 - 4 M^2\right)^2 \xi^2}.
	\end{align}
	It can be checked that for a given value of $N_0$, $M$ and $\xi$, Eqs. \eqref{kxi1}, \eqref{h0}, \eqref{ns_analytic} and \eqref{r_analytic} provide a valid approximation of the numerical results displayed in Figs. \ref{fig:ns_r_k}-\ref{fig:ns_dns_n0}. For $77 \lesssim |\xi| \lesssim 24000$ and $M \simeq M_{\text{GUT}} = 2\times 10^{16}$ GeV, we obtain $0.0034 \lesssim \kappa \lesssim 1$, $0.9637 \lesssim n_s \lesssim 0.965$, $0.0035 \lesssim r \lesssim 0.0037$ and $dn_s/d \ln k \simeq -(1.26 - 1.20) \times 10^{-3}$. The large tensor modes predicted by the model are potentially measurable by forthcoming CMB experiments such as, LiteBIRD \cite{LiteBIRD:2020khw}, Simons Observatory \cite{SimonsObservatory:2018koc}, PRISM \cite{Andre:2013afa}, CMB-S4 \cite{CMB-S4:2020lpa}, CMB-HD \cite{Sehgal:2019ewc}, CORE \cite{Finelli:2016cyd} and PIXIE \cite{Kogut:2011xw}.
	%In the large $\xi$ limit with $|\xi|M^2 \ll 1$, we obtain \cite{Bezrukov:2007ep,Okada:2010jf} the following approximate expressions for $h_0$, $h_e$, $n_s$, $r$ and $dn_s/d \ln k$,
	%\begin{align}
	%	h_0 &\simeq \sqrt{\frac{4N_0}{3 \,\xi}}, \qquad  h_e \simeq \left( \frac{4}{3} \right)^{1/4} \frac{1}{\sqrt{\xi}} ,\\
	%	n_s &\simeq 1 - \frac{2}{N_0}, \qquad r \simeq \frac{12}{N_0^2}, \\
	%	\dfrac{dn_s}{d \ln k} &\simeq  -\frac{r(1-n_s)}{2}\simeq -\frac{12}{N_0^3}.
	%\end{align}
	%It can readily be checked that for $N_0$ in the range $53.1 - 55$, $n_s$, $r$ and $dn_s/d \ln k$ vary in the ranges $$
	\section{\large{\bf Metastable Cosmic string and Radiative Breaking of $U(1)_{\chi}$} Symmetry}\label{sec7}
	
	In this section, we study the SGWB spectra produced by the decay of a metastable cosmic string network \cite{Ellis:2020ena,King:2020hyd}. The embedding of the $SU(5)\times U(1)_\chi$ group in the $SO(10)$ group leads to the production of a metastable cosmic string network, when the $U(1)_\chi$ symmetry is broken by non-zero VEV of $\mathbf{16}$ representation of $SO(10)$. This metastable cosmic string network can decay via the Schwinger production of monopole-antimonopole pairs which generates a stochastic gravitational wave background (SGWB) within the frequency range of ongoing and future gravitational wave (GW) experiments. For the case of stable cosmic strings production in $SU(5)\times U(1)_\chi$ model, see ref \cite{ Ahmed:2022vlc, Ahmed:2022thr}.
	
	As shown in Eq. \eqref{eq:breaking_pattern}, the breaking of $SO(10)$ into $G_{\text{SM}}$ is achieved in three steps through $SU(5) \times U(1)_\chi$. First, the group $SO(10)$ breaks to $SU(5) \times U(1)_\chi$ by the non-zero VEV of $\mathbf{45}_H$ in $(\mathbf{1}, 0)$ direction. Subsequently, in the second step, $SU(5) \times U(1)_\chi$ breaks to $G_{\text{SM}} \times U(1)_\chi$ due to the non-zero VEV of the $\Phi(\mathbf{24}, 0)$ representation in the hypercharge direction $\Phi_{24} (\mathbf{1}, \mathbf{1}, 0)$. Finally, the $U(1)_{\chi}$ symmetry breaks via the non-zero VEV of $\mathbf{16}_H$ representation in the right-handed neutrino direction $\nu_H(\mathbf{1}, -5)$. During the first and second breaking steps, magnetic monopoles are formed. However, in the final step, metastable cosmic strings are generated, as the homotopy group of the manifold $\mathcal{M} = SO(10)/G_{\text{SM}}$ is trivial, $\pi_1(\mathcal{M}) = I$. 
	
	The radiative breaking of $U(1)_{\chi}$ is achieved as follows. After the end of inflation, the effective unbroken gauge symmetry is $SU(3)_C \times SU(2)_L \times U(1)_Y \times U(1)_{\chi}$. The superpotential terms relevant for $U(1)_{\chi}$ symmetry breaking are given by
	
	\begin{equation}
		\label{W_u1_breaking}
		W = W_{\text{MSSM}} + \sigma_{\chi} S \nu_H \bar{\nu}_H. 
	\end{equation}
	From the above equation, we obtain
	\begin{eqnarray} 
		F_S^{\dagger} &=& \frac{\partial W}{\partial S} = \sigma_{\chi} \nu_H \bar{\nu}_H = 0 ,\nonumber\\ 
		F_{\bar{\nu}_H}^{\dagger} &=& \frac{\partial W}{\partial \bar{\nu}_H} = \sigma_{\chi} S \nu_H   = 0 ,\nonumber\\ 
		F_{\nu_H}^{\dagger} &=& \frac{\partial W}{\partial \nu_H} = \sigma_{\chi} S \bar{\nu}_H.
	\end{eqnarray}
	This leads to the following vacua;
	\begin{equation}
		\langle \nu_H \rangle = \langle \bar{\nu}_H \rangle= 0, \qquad  \langle S \rangle = \text{Arbitrary}.
	\end{equation}
	In order to break the $U(1)_{\chi}$ symmetry, a non-zero VEV of the fields $\nu_H, \bar{\nu}_H$ is desired; $\langle \nu_H \rangle = \langle \bar{\nu}_H \rangle \neq 0$.  Including the soft SUSY breaking mass terms,
	\begin{eqnarray} 
		V_{\text{Soft}} &=& m_S^2 \vert S \vert^2 + m_{\nu_H}^2 \vert \nu_H \vert^2 + m_{\bar{\nu}_H}^2 \vert \bar{\nu}_H \vert^2  \nonumber\\
		&+&  A_{\chi} \sigma_{\chi} S \nu_H \bar{\nu}_H + \frac{1}{2} M_{\chi} Z_{\chi} Z_{\chi} , 
	\end{eqnarray}
	where $A_{\chi}$ is the coefficient of linear soft mass terms, $Z_{\chi}$ is the $U(1)_{\chi}$ gaugino and $M_{\chi}$ is the gaugino mass. The full scalar potential is then given by,
	\begin{eqnarray}
		\nonumber
		V &=& V_{F} + V_{D} + V_{\text{Soft}} \nonumber\\ 
		&=&  \sigma_{\chi}^2 \left(\lvert \nu_H \rvert^2 \lvert \bar{\nu}_H \rvert^2 +  \lvert S \rvert^2 \lvert \bar{\nu}_H  \rvert^2 + \lvert S \rvert^2 \lvert \nu_H  \rvert^2\right) 
		+ \frac{25}{2} g_{\chi}^2 \left( \vert \nu_H \vert^2 - \vert \bar{\nu}_H \vert^2 \right)^2 \nonumber\\ 
		&+& m_S^2 \vert S \vert^2 + m_{\nu_H}^2 \vert \nu_H  \vert^2 + m_{\bar{\nu}_H}^2 \vert \bar{\nu}_H \vert^2  + A_{\chi} \sigma_{\chi} S \nu_H \bar{\nu}_H + \frac{1}{2} M_{\chi} Z_{\chi} Z_{\chi} .
	\end{eqnarray}
	The potential minima can be obtained in the $D$-flat direction ($\lvert \nu_H  \rvert = \lvert \bar{\nu}_H  \rvert$) as follows;
	\begin{eqnarray} 
		\frac{\partial V}{\partial S^{\dagger}} &=& \left(2 \sigma_{\chi}^2  \vert \nu_H \vert^2 + m_{S}^2\right) S = 0 ,\nonumber\\ 
		\frac{\partial V}{\partial \nu_H^{\dagger}}&=& \left(\sigma_{\chi}^2  \left( \vert \nu_H \vert^2 + \vert S \vert^2\right) +  m_{\nu_H}^2\right)  \nu_H= 0 .
	\end{eqnarray}
	The VEV of the fields $\nu_H$, $\bar{\nu}_H$ and $S$ is found to be,
	\begin{equation}
		\langle \vert \nu_H \vert\rangle=\langle \vert \bar{\nu}_H \vert\rangle \simeq \sqrt{-\frac{m_S^2}{2 \, \sigma_{\chi}^2}}\ , \qquad \langle S \rangle \simeq \sqrt{-\frac{m_{\nu_H}^2}{ \sigma_{\chi}^2}}\ . 
	\end{equation}
	The negative mass squared, $m_{S}^2 < 0$ should be satisfied at an intermediate scale $m_{\chi}$ below the GUT scale to realize the correct $U(1)_{\chi}$ symmetry breaking. A negative mass squared can be achieved through the RG running from the GUT scale to an intermediate scale with a large enough Yukawa coupling even if the mass squared is positive at the GUT scale. 
	We consider the $U(1)_{\chi}$ renormalization group equations and analyze the running of the scalar masses $m_{\nu_H}^2$ and $m_{S}^2$. A negative mass-squared $m_{S}^2$ will trigger the radiative breaking of $U(1)_{\chi}$ symmetry. We show that the mass-squared of the fields $\nu_H$ and $S$ evolve in such a way that $m_S^2$ becomes negative whereas, $m_{\nu_H}^2$ remain positive. 
	
	The renormalization group equations are given by
	%for the soft SUSY breaking scalar masses of $U(1)_{\chi}$ Higgs field $\chi$ and the right handed sneutrino $\nu^c$ are given by
	\begin{eqnarray} 
		16 \pi^2 \frac{d g_{\chi}}{d t} &=& \frac{153}{5} g_{\chi}^3 , \\
		16 \pi^2 \frac{d M_{\chi}}{d t} &=& \frac{153}{10} g_{\chi}^2 M_{\chi} , \\
		16 \pi^2 \frac{d \sigma_{\chi}}{d t} &=& \frac{\sigma_{\chi}}{2} \left( 6 \sigma_{\chi}^2 - 5 g_{\chi}^2  \right) , \\
		16 \pi^2 \frac{d m_{\nu_H}^2}{d t} &=&  2 \sigma_{\chi}^2 \left( m_S^2 + 2m_{\nu_H}^2   \right)  + 2 T_{\sigma_{\chi}}^2  - 5 g_{\chi}^2 M_{\chi}^2 ,  \\ 
		16 \pi^2 \frac{d m_{S}^2}{d t} &=& 2 \sigma_{\chi}^2 \left( m_S^2 + 2m_{\nu_H}^2   \right) +  T_{\sigma_{\chi}}^2 , \\
		16 \pi^2 \frac{d T_{\sigma_{\chi}}}{d t} &=& T_{\sigma_{\chi}} \left( 9 \sigma_{\chi}^2 -\frac{5}{2} g_{\chi}^2 \right) + 5 \sigma_{\chi}  g_{\chi}^2 M_{\chi},
	\end{eqnarray}
	\begin{figure}[t]
		\centering \includegraphics[width=8.035cm]{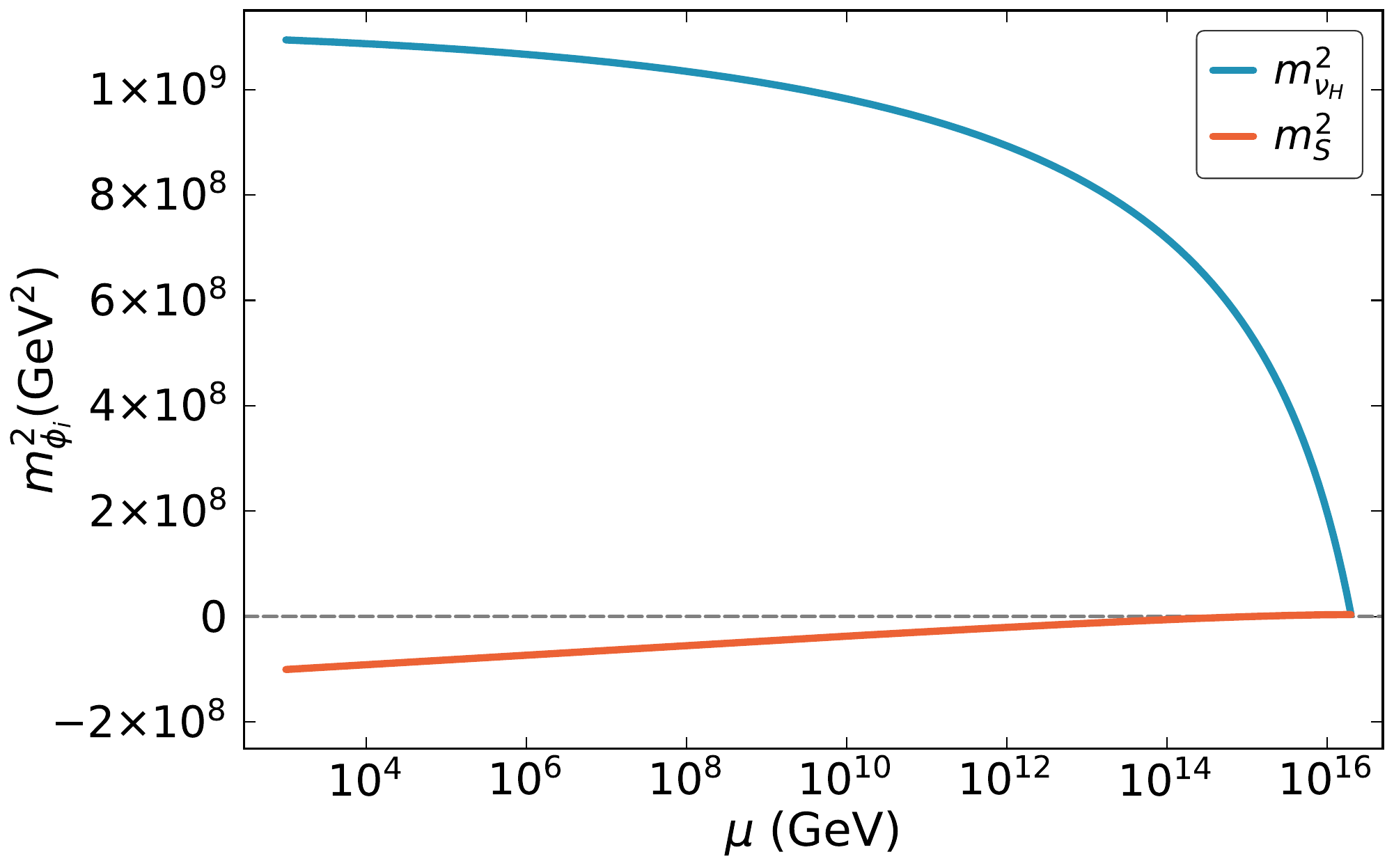}
		\centering \includegraphics[width=8.035cm]{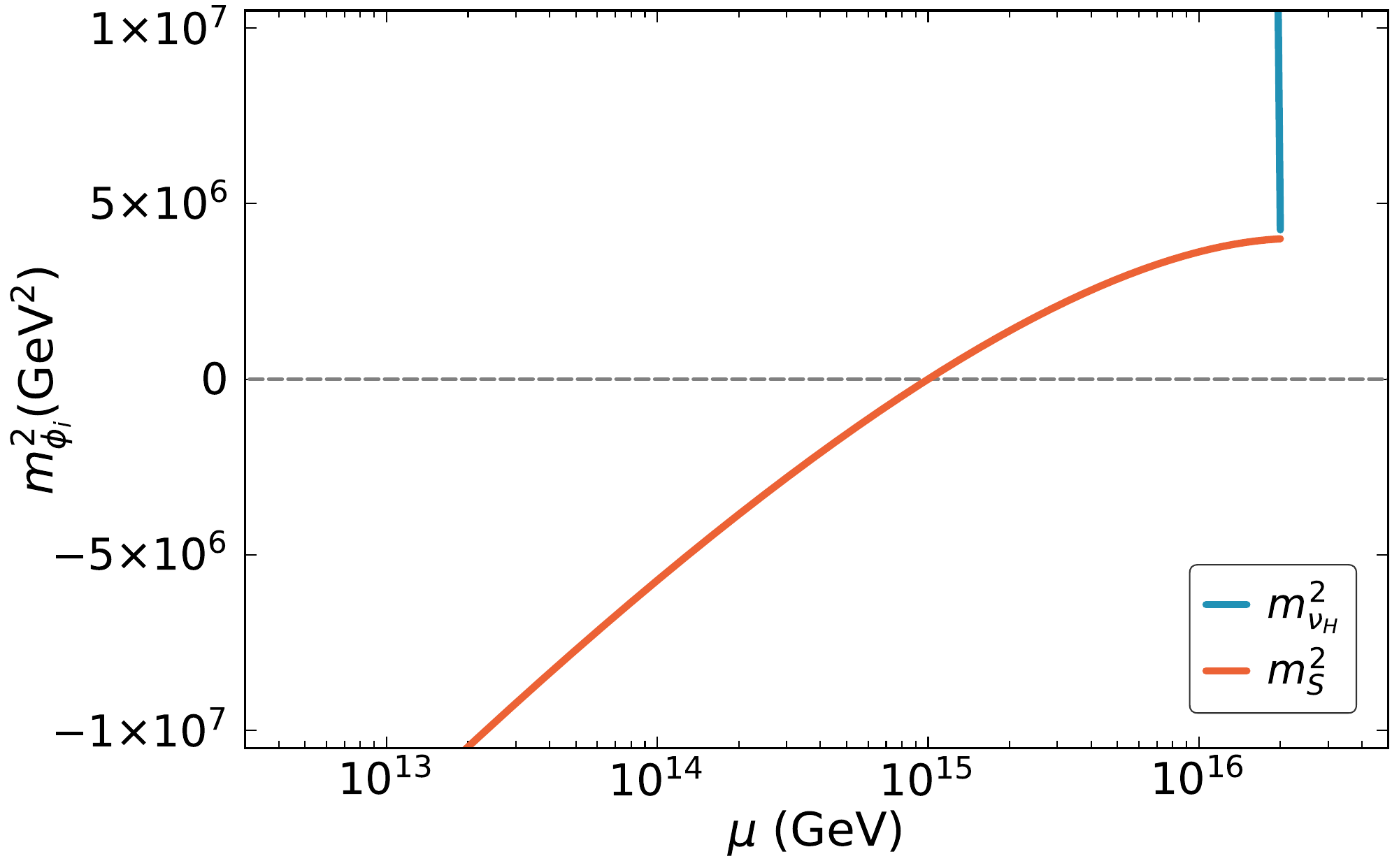}
		\caption{The evolution of scalar squared masses $m_{\nu_H}^2$ and $m_S^2$ from the GUT to TeV scale. The mass-squared $m_{S}^2$ quickly becomes negative at a scale of $m_{\chi} \simeq 10^{15}$ GeV, while $m_{\nu_H}^2$ exhibits rapid growth and approaches around $10^9 \text{ GeV}^2$ at TeV scale.}
		\label{RGEs}
	\end{figure}
	where $t = \ln \mu$. The evolution of these parameters depends on the boundary conditions at GUT scale, $M_{\text{GUT}}=2\times 10^{16}$ GeV.  We assume universal soft SUSY breaking at this scale,

	\begin{gather} \label{mzb}
		\quad m_{\nu_H}^2 = m_{S}^2 = m_0^2, \quad M_{\chi} =10^5 \text{ TeV}, \nonumber \\ 
		g_{\chi}^2=1.04,\quad \sigma_{\chi} = 0.3783, \quad m_0 = 2 \text{ TeV}, \\ \nonumber
		\quad T_{\sigma_{\chi}} = \sigma_{\chi} A_{\chi}= 5 \text{ TeV}.
	\end{gather}
	As shown in Fig. \ref{RGEs}, the mass-squared $m_{S}^2$ quickly becomes negative at a scale of $m_{\chi} \simeq 10^{15}$ GeV, while $m_{\nu_H}^2$ exhibits rapid growth and approaches around $10^9 ~ \text{GeV}^2$ near the TeV scale.
	
	After inflation, the breaking of $U(1){\chi}$ leads to the formation of topologically metastable cosmic strings. These strings are subject to observational constraints, which are expressed using the dimensionless quantity $G_N \mu_s$, representing the gravitational interaction strength of the strings. Here, $G_N$ denotes Newton's constant, and $\mu_s \simeq 2 \pi m_{\chi}^2$ represents the mass per unit length of the string, with $m_{\chi}$ being the $U(1)_{\chi}$ breaking scale. The Planck bound on $G_N \mu_s$, obtained from constraints on the string's contribution to the cosmic microwave background (CMB) power spectrum, is given by \cite{Planck:2018jri, Planck:2018vyg}	
	\begin{equation}
		G_N \mu_s \lesssim 2.4 \times 10^{-7}.
	\end{equation}
	Consequently, an upper bound on the $U(1)_{\chi}$ breaking scale can be determined:
	\begin{equation}
		m_{\chi} \lesssim 2.35 \times 10^{15} ~ \text{GeV},
	\end{equation}
	which is easily obtained as evident from Fig. \ref{RGEs}, and depends on the initial boundary conditions at GUT scale. As the strings form after inflation, the stochastic gravitational wave background is generated from undiluted strings, as discussed in the following section.
	
	\subsection{Gravitational Waves From Cosmic Strings}

The combined effects of various sources of gravitational waves (GW), such as those stemming from inflation, cosmic strings, and phase transitions, lead to the formation of stochastic gravitational wave background (SGWB). Among these, the GWs arising from tensor perturbations upon re-entry into the cosmic horizon are responsible for the inflationary SGWB. However, the inflationary SGWB abundance ($\Omega_{\text{GW}}h^2\sim 10^{-16}$) is far too small to be detectable by PTA experiments. The detection of these gravitational waves necessitates specialized approaches, particularly involving the identification of tensor modes, which are briefly discussed in Sec 3. 
 
	The radiative breaking of the $U(1)_{\chi}$ gauge symmetry gives rise to the emergence of a metastable cosmic string network. This network has the potential to undergo decay via the Schwinger production of monopole-antimonopole pairs, consequently leading to the generation of a  SGWB. The SGWB arising from metastable cosmic string network are expressed relative to critical density as \cite{Blanco-Pillado:2017oxo}
	\begin{align}
		\Omega_\text{GW}(f) = \frac{\partial \rho_\text{gw}(f)}{\rho_c \partial \ln f}= \frac{8 \pi f (G \mu_{CS})^2}{3 H_0^2} \sum_{n = 1}^\infty C_n(f) \, P_n \,,
		\label{eq:Omega}
	\end{align}
	where $\rho_\text{gw}$ denotes the GW energy density, $\rho_c$ is the critical energy density of the universe, and $H_0 = 100 \,h \textrm{ km}\textrm{s}^{-1} \textrm{ Mpc}^{-1}$ is the Hubble parameter. The parameter $P_n \simeq\frac{50}{\zeta(4/3)n^{4/3}}$ \cite{Auclair:2019wcv} is the power spectrum of GWs emitted by the $n^{\rm th}$ harmonic of a cosmic string loop and $C_n(f)$ indicates the number of loops emitting GWs that are observed at a given frequency $f$
	\begin{align}
		\label{eq:Cn}
		C_n(f) = \frac{2 n}{f^2} \int_{z_\text{min}}^{z_\text{max}}dz\:\frac{\mathcal{N}\left(\ell\left(z\right),\,t\left(z\right)\right)}{H\left(z\right)(1 + z)^6} \,,
	\end{align}
	which is a function of number density of cosmic string loops $\mathcal{N}(\ell,t)$, with $\ell = 2n/((1 + z) f)$. For the number density of cosmic string loops, $\mathcal{N}(\ell,t)$, we use the approximate expressions of Blanco-Pillado-Olum-Shlaer (BOS) model from \cite{Blanco-Pillado:2017oxo,Auclair:2019wcv}
	\begin{align}
		\mathcal{N}_r(\ell,t)  &= \frac{0.18}{t^{3/2}(\ell+\Gamma G\mu_{s} t)^{5/2}},\label{eq:nr}\\
		\mathcal{N}_{m,r}(\ell,t)  &= \frac{0.18\sqrt{t_{eq}}}{t^2(\ell+\Gamma G\mu_{s} t)^{5/2}}
		=\frac{0.18(2H_0\sqrt{\Omega_r})^{3/2}}{(\ell+\Gamma G\mu_{s} t)^{5/2}}(1+z)^3~.
	\end{align}
	For our region of interest, the dominant contribution is obtained from the loops generated during the radiation-dominated era. For $t(z)$ and $H(z)$, we use the expressions for $\Lambda$CDM cosmology assuming a standard thermal history of universe, while ignoring the changes in the number of effective degrees of freedom with $z$ 
	\begin{eqnarray}\label{OmegaGW}
		H(z)&=&H_0\sqrt{\Omega_\Lambda + \Omega_m(1+z)^3+\Omega_r(1+z)^4},\\
		t(z) &=& \int_{z_\text{min}}^{z_{\text{max}}} \frac{dz' }{H(z')(1+z')},\quad l(z)=\frac{2n}{(1+z)f}. 
		%,\\H(z)&=&H_0 (1+z)^2\sqrt{\Omega_r}, \quad t(z)=\frac{1}{2(1+z)^2H_0\sqrt{\Omega_r}}
	\end{eqnarray}
	\begin{figure}[t]
		\centering \includegraphics[width=9.00cm]{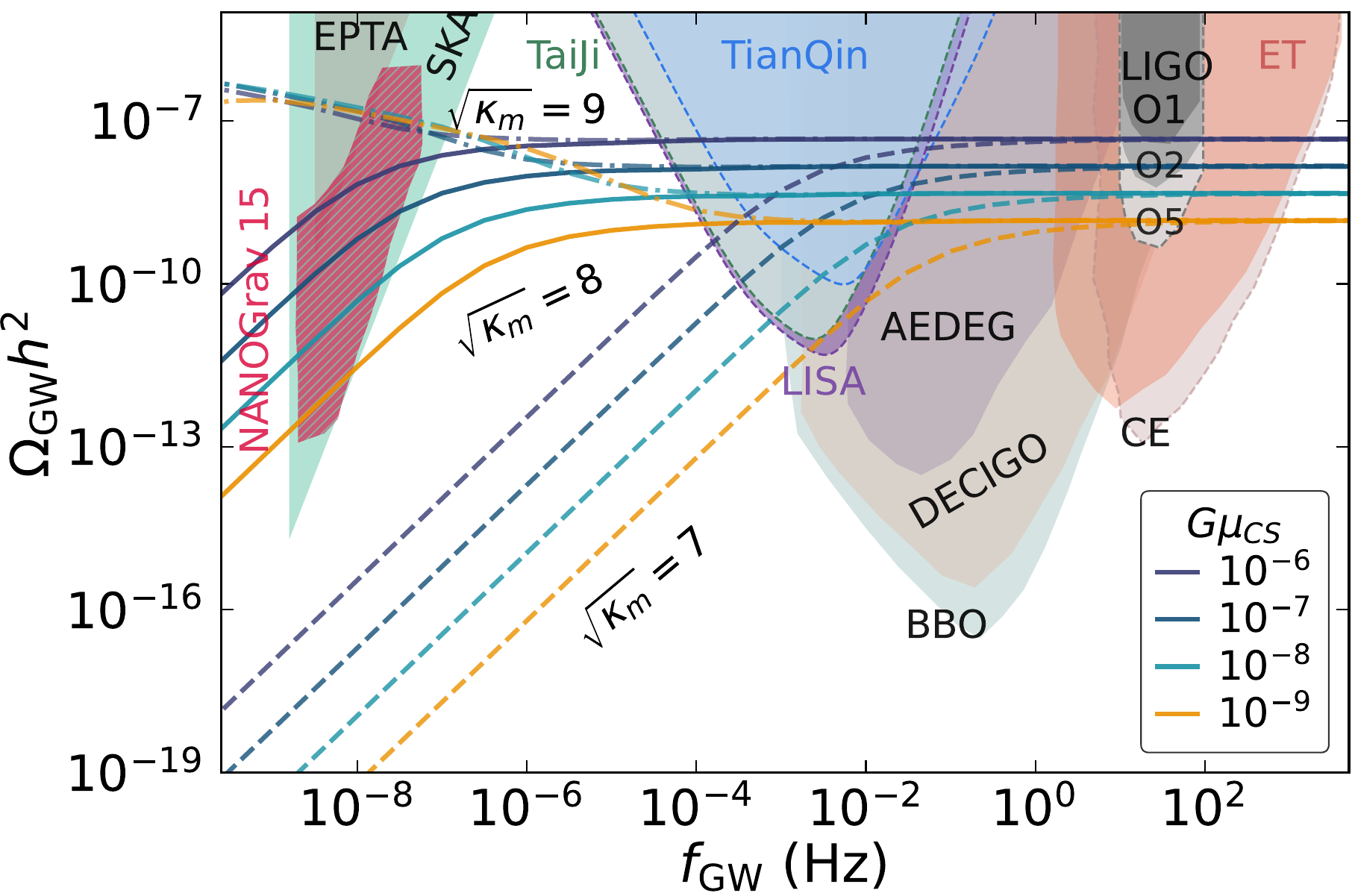}
		\caption{Gravitational wave spectra from metastable cosmic strings explaining the NANOGrav excess at 2-$\sigma$ confidence level. The curves are drawn for string tension $10^{-9} \lesssim G\mu_{s} \lesssim 10^{-6}$, with $\sqrt{\kappa_m}=7,8,9$. The shaded areas in the background indicate the sensitivities of the current and future experiments.}
		\label{omegagw}
	\end{figure}
	The integration range in the above equation corresponds to the lifetime of the cosmic string network, from its formation at $z_\text{max} \simeq \frac{T_r}{2.7K}$ until its decay at $z_\text{min}$  given by \cite{Leblond:2009fq,Monin:2008mp,KaiSchmitz}, 
	\begin{equation}
		z_\text{min} = \left( \frac{70}{H_0}\right)^{1/2} \left( \Gamma  \; \Gamma_d  \; G \mu_{s} \right)^{1/4},\quad \Gamma_d =\frac{\mu_s}{2\pi}e^{-\pi\kappa_m}, \quad \kappa_m = \frac{m_M^2}{\mu_s},
		\label{zmin}
	\end{equation}
	where $m_M$ represents the monopole mass, and $\mu_s$ is the string tension. In numerical calculation, we take the value of $\Gamma \simeq 50$ and the reheat temperature is fixed at $\sim 10^9$ GeV. The dimensionless parameter $\kappa_m$ quantifies the hierarchy between the GUT and $U(1)_{\chi}$ breaking scales.
	
	Figure \ref{omegagw} displays the gravitational wave spectra resulting from metastable cosmic strings, for the string tension $10^{-9} \lesssim G\mu_{s} \lesssim 10^{-6}$. The three sets of curves are drawn for specific ratios of the GUT and $U(1)_{\chi}$ breaking scales, namely, $\sqrt{\kappa_m}=7, 8, 9$. Note that both for $\sqrt{\kappa_m}=8, 9$, the spectra passes through the sensitivity bounds of NANOGrav 15-years data bounds. It is evident from the  Fig. \ref{omegagw} that the entire range of $G\mu_{s}$ produces gravitational wave spectra that can be detected by most gravitational wave detectors. For instance, LIGO O1 \cite{LIGOScientific:2019vic} has already ruled out cosmic string formation at $G\mu_{CS} \lesssim 10^{-6}$ in the high-frequency regime of $10$-$100$ Hz. On the other hand, the low-frequency band of $1$-$10$ nHz can be explored by NANOGrav \cite{NANOGrav:2023hvm, NANOGrav:2023gor}, EPTA \cite{Ferdman:2010xq}, and other gravitational wave experiments operating at nano Hz frequencies.
	
	Higher frequency gravitational wave spectra can be probed by pulsar timing arrays such as SKA \cite{Smits:2008cf}, space-based laser interferometers like LISA \cite{LISA:2017pwj}, Taiji \cite{Hu:2017mde}, TianQin \cite{TianQin:2015yph}, BBO \cite{Corbin:2005ny}, DECIGO \cite{Seto:2001qf}, and ground-based interferometers like Einstein Telescope \cite{Punturo:2010zz} (ET), Cosmic Explorer \cite{LIGOScientific:2016wof} (CE), and atomic interferometer AEDGE \cite{AEDGE:2019nxb}. All these instruments have the potential to explore gravitational waves from metastable cosmic strings across a wide regime of frequencies.

	\section{\large{\bf Summary}}\label{sec8}
	We have explored a realistic GUT framework involving $SU(5) \times U(1)_{\chi}$ group embedded within the larger $SO(10)$ group. The breaking of $SO(10)$ to the $G_{\text{SM}}$ group is achieved through an intermediate step of $SU(5) \times U(1)_{\chi}$, via the non-zero vacuum expectation values (VEVs) of the $\mathbf{45}_H$, $\mathbf{16}_H$, and $\widebar{\mathbf{16}}_H$ Higgs representations of $SO(10)$. This framework also incorporates a viable model of non-minimal Higgs inflation, in which the $SU(5)$ gauge symmetry is spontaneously broken during inflation, effectively resolving the monopole problem. The model yields a scalar tilt that is consistent with the Planck 2018 bounds, along with large tensor modes, potentially measurable by forthcoming CMB experiments. 
	
	Subsequently, after the end of inflation, the radiative breaking of the $U(1)_{\chi}$ symmetry at a slightly lower scale $m_{\chi} < M_{\text{GUT}}$ leads to the formation of a metastable cosmic string network. This metastable string network can decay via the Schwinger production of monopole-antimonopole pairs, generating a stochastic gravitational wave background that is compatible with the most recent pulsar timing array (PTA) experiments, such as NANOGrav, CPTA, EPTA and PPTA. Moreover, this gravitational wave spectrum is accessible to both existing and future ground-based and space-based experiments.

	%%%%%%%%%%%%%%%%%%%%%%%%%%%%%%%%%
	
	%%%%%%%%%%%%%%%%%%%%%%%%%%%%%%%


\begin{thebibliography}{99}


 
			%\cite{NANOGrav:2023hvm}
		\bibitem{NANOGrav:2023hvm}
		A.~Afzal \textit{et al.} [NANOGrav],
		%``The NANOGrav 15 yr Data Set: Search for Signals from New Physics,''
		Astrophys. J. Lett. \textbf{951}, no.1, L11 (2023)
	%	doi:10.3847/2041-8213/acdc91
		[arXiv:2306.16219 [astro-ph.HE]].
		%87 citations counted in INSPIRE as of 16 Jul 2023	
		%\cite{NANOGrav:2023gor}
		\bibitem{NANOGrav:2023gor}
		G.~Agazie \textit{et al.} [NANOGrav],
		%``The NANOGrav 15 yr Data Set: Evidence for a Gravitational-wave Background,''
		Astrophys. J. Lett. \textbf{951}, no.1, L8 (2023)
	%	doi:10.3847/2041-8213/acdac6
		[arXiv:2306.16213 [astro-ph.HE]].
		%91 citations counted in INSPIRE as of 16 Jul 2023


  
		
		
		%\cite{Antoniadis:2023rey}
		\bibitem{Antoniadis:2023rey}
		J.~Antoniadis, P.~Arumugam, S.~Arumugam, S.~Babak, M.~Bagchi, A.~S.~B.~Nielsen, C.~G.~Bassa, A.~Bathula, A.~Berthereau and M.~Bonetti, \textit{et al.}
		%``The second data release from the European Pulsar Timing Array III. Search for gravitational wave signals,''
		[arXiv:2306.16214 [astro-ph.HE]].
		%75 citations counted in INSPIRE as of 16 Jul 2023
		
		%\cite{Reardon:2023gzh}
		\bibitem{Reardon:2023gzh}
		D.~J.~Reardon, A.~Zic, R.~M.~Shannon, G.~B.~Hobbs, M.~Bailes, V.~Di Marco, A.~Kapur, A.~F.~Rogers, E.~Thrane and J.~Askew, \textit{et al.}
		%``Search for an Isotropic Gravitational-wave Background with the Parkes Pulsar Timing Array,''
		Astrophys. J. Lett. \textbf{951}, no.1, L6 (2023)
	%	doi:10.3847/2041-8213/acdd02
		[arXiv:2306.16215 [astro-ph.HE]].
		%74 citations counted in INSPIRE as of 16 Jul 2023
		
		%\cite{Xu:2023wog}
		\bibitem{Xu:2023wog}
		H.~Xu, S.~Chen, Y.~Guo, J.~Jiang, B.~Wang, J.~Xu, Z.~Xue, R.~N.~Caballero, J.~Yuan and Y.~Xu, \textit{et al.}
		%``Searching for the Nano-Hertz Stochastic Gravitational Wave Background with the Chinese Pulsar Timing Array Data Release I,''
		Res. Astron. Astrophys. \textbf{23}, no.7, 075024 (2023)
	%	doi:10.1088/1674-4527/acdfa5
		[arXiv:2306.16216 [astro-ph.HE]].
		%76 citations counted in INSPIRE as of 16 Jul 2023
		
		
		%\cite{NANOGrav:2023hde}
		\bibitem{NANOGrav:2023hde}
		G.~Agazie \textit{et al.} [NANOGrav],
		%``The NANOGrav 15 yr Data Set: Observations and Timing of 68 Millisecond Pulsars,''
		Astrophys. J. Lett. \textbf{951}, no.1, L9 (2023)
		%doi:10.3847/2041-8213/acda9a
		[arXiv:2306.16217 [astro-ph.HE]].
		%30 citations counted in INSPIRE as of 16 Jul 2023
		
		
		%\cite{NANOGrav:2020bcs}
\bibitem{NANOGrav:2020bcs}
Z.~Arzoumanian \textit{et al.} [NANOGrav],
%``The NANOGrav 12.5 yr Data Set: Search for an Isotropic Stochastic Gravitational-wave Background,''
Astrophys. J. Lett. \textbf{905}, no.2, L34 (2020)
%doi:10.3847/2041-8213/abd401
[arXiv:2009.04496 [astro-ph.HE]].
%700 citations counted in INSPIRE as of 14 Aug 2023

		
		
		
		%\cite{Antoniadis:2023zhi}
		\bibitem{Antoniadis:2023zhi}
		J.~Antoniadis, P.~Arumugam, S.~Arumugam, P.~Auclair, S.~Babak, M.~Bagchi, A.~S.~Bak Nielsen, E.~Barausse, C.~G.~Bassa and A.~Bathula, \textit{et al.}
		%``The second data release from the European Pulsar Timing Array: V. Implications for massive black holes, dark matter and the early Universe,''
		[arXiv:2306.16227 [astro-ph.CO]].
		%34 citations counted in INSPIRE as of 16 Jul 2023
		
		
		
		%\cite{NANOGrav:2023hfp}
		\bibitem{NANOGrav:2023hfp}
		G.~Agazie \textit{et al.} [NANOGrav],
		%``The NANOGrav 15-year Data Set: Constraints on Supermassive Black Hole Binaries from the Gravitational Wave Background,''
		[arXiv:2306.16220 [astro-ph.HE]].
		%37 citations counted in INSPIRE as of 16 Jul 2023
		
		
	
		
		
		
		%\cite{Goncharov:2021oub}
		\bibitem{Goncharov:2021oub}
		B.~Goncharov, R.~M.~Shannon, D.~J.~Reardon, G.~Hobbs, A.~Zic, M.~Bailes, M.~Curylo, S.~Dai, M.~Kerr and M.~E.~Lower, \textit{et al.}
		%``On the Evidence for a Common-spectrum Process in the Search for the Nanohertz Gravitational-wave Background with the Parkes Pulsar Timing Array,''
		Astrophys. J. Lett. \textbf{917}, no.2, L19 (2021)
	%	doi:10.3847/2041-8213/ac17f4
		[arXiv:2107.12112 [astro-ph.HE]].
		%247 citations counted in INSPIRE as of 16 Jul 2023
		
		
		
		
		%\cite{Chen:2021rqp}
		\bibitem{Chen:2021rqp}
		S.~Chen, R.~N.~Caballero, Y.~J.~Guo, A.~Chalumeau, K.~Liu, G.~Shaifullah, K.~J.~Lee, S.~Babak, G.~Desvignes and A.~Parthasarathy, \textit{et al.}
		%``Common-red-signal analysis with 24-yr high-precision timing of the European Pulsar Timing Array: inferences in the stochastic gravitational-wave background search,''
		Mon. Not. Roy. Astron. Soc. \textbf{508}, no.4, 4970-4993 (2021)
	%	doi:10.1093/mnras/stab2833
		[arXiv:2110.13184 [astro-ph.HE]].
		%211 citations counted in INSPIRE as of 16 Jul 2023
		
		
		
		
		%\cite{Kibble:1976sj}
		\bibitem{Kibble:1976sj}
		T.~W.~B.~Kibble,
		%``Topology of Cosmic Domains and Strings,''
		J. Phys. A \textbf{9}, 1387-1398 (1976)
	%	doi:10.1088/0305-4470/9/8/029
		%3002 citations counted in INSPIRE as of 16 Jul 2023
		
		
		
		%\cite{Hindmarsh:1994re}
		\bibitem{Hindmarsh:1994re}
		M.~B.~Hindmarsh and T.~W.~B.~Kibble,
		%``Cosmic strings,''
		Rept. Prog. Phys. \textbf{58}, 477-562 (1995)
	%	doi:10.1088/0034-4885/58/5/001
		[arXiv:hep-ph/9411342 [hep-ph]].
		%1092 citations counted in INSPIRE as of 16 Jul 2023

  

  %\cite{Buchmuller:2023aus}
\bibitem{Buchmuller:2023aus}
W.~Buchmuller, V.~Domcke and K.~Schmitz,
%``Metastable cosmic strings,''
[arXiv:2307.04691 [hep-ph]].
%6 citations counted in INSPIRE as of 17 Aug 2023

%\cite{Antusch:2023zjk}
\bibitem{Antusch:2023zjk}
S.~Antusch, K.~Hinze, S.~Saad and J.~Steiner,
%``Singling out SO(10) GUT models using recent PTA results,''
[arXiv:2307.04595 [hep-ph]].
%10 citations counted in INSPIRE as of 17 Aug 2023

%\cite{Fu:2023mdu}
\bibitem{Fu:2023mdu}
B.~Fu, S.~F.~King, L.~Marsili, S.~Pascoli, J.~Turner and Y.~L.~Zhou,
%``Testing Realistic $SO(10)$ SUSY GUTs with Proton Decay and Gravitational Waves,''
[arXiv:2308.05799 [hep-ph]].
%1 citations counted in INSPIRE as of 17 Aug 2023

   %\cite{Lazarides:2023rqf}
\bibitem{Lazarides:2023rqf}
G.~Lazarides, R.~Maji, A.~Moursy and Q.~Shafi,
%``Inflation, superheavy metastable strings and gravitational waves in non-supersymmetric flipped SU(5),''
[arXiv:2308.07094 [hep-ph]].
%0 citations counted in INSPIRE as of 17 Aug 2023

  
		%\cite{Ellis:2020ena}
		\bibitem{Ellis:2020ena}
		J.~Ellis and M.~Lewicki,
		%``Cosmic String Interpretation of NANOGrav Pulsar Timing Data,''
		Phys. Rev. Lett. \textbf{126}, no.4, 041304 (2021)
	%	doi:10.1103/PhysRevLett.126.041304
		[arXiv:2009.06555 [astro-ph.CO]].
		%192 citations counted in INSPIRE as of 16 Jul 2023
		
		%\cite{King:2020hyd}
		\bibitem{King:2020hyd}
		S.~F.~King, S.~Pascoli, J.~Turner and Y.~L.~Zhou,
		%``Gravitational Waves and Proton Decay: Complementary Windows into Grand Unified Theories,''
	%	Phys. Rev. Lett. \textbf{126}, no.2, 021802 (2021)
		%doi:10.1103/PhysRevLett.126.021802
		[arXiv:2005.13549 [hep-ph]].
		%54 citations counted in INSPIRE as of 16 Jul 2023
		
		
		
		
		
		%\cite{Buchmuller:2020lbh}
		\bibitem{Buchmuller:2020lbh}
		W.~Buchmuller, V.~Domcke and K.~Schmitz,
		%``From NANOGrav to LIGO with metastable cosmic strings,''
		Phys. Lett. B \textbf{811}, 135914 (2020)
	%	doi:10.1016/j.physletb.2020.135914
		[arXiv:2009.10649 [astro-ph.CO]].
		%104 citations counted in INSPIRE as of 16 Jul 2023
		
		
		%\cite{Ahmed:2021ucx}
		\bibitem{Ahmed:2021ucx}
		W.~Ahmed, M.~Junaid and U.~Zubair,
		%``Primordial black holes and gravitational waves in hybrid inflation with chaotic potentials,''
		Nucl. Phys. B \textbf{984}, 115968 (2022)
		%doi:10.1016/j.nuclphysb.2022.115968
		[arXiv:2109.14838 [astro-ph.CO]].
		%24 citations counted in INSPIRE as of 16 Jul 2023
		
		
		
		%\cite{Afzal:2022vjx}
		\bibitem{Afzal:2022vjx}
		A.~Afzal, W.~Ahmed, M.~U.~Rehman and Q.~Shafi,
		%``\ensuremath{\mu}-hybrid inflation, gravitino dark matter, and stochastic gravitational wave background from cosmic strings,''
		Phys. Rev. D \textbf{105}, no.10, 103539 (2022)
		%doi:10.1103/PhysRevD.105.103539
		[arXiv:2202.07386 [hep-ph]].
		%15 citations counted in INSPIRE as of 16 Jul 2023
		
		
		
		%\cite{Vagnozzi:2023lwo}
		\bibitem{Vagnozzi:2023lwo}
		S.~Vagnozzi,
		%``Inflationary interpretation of the stochastic gravitational wave background signal detected by pulsar timing array experiments,''
		%doi:10.1016/j.jheap.2023.07.001
		[arXiv:2306.16912 [astro-ph.CO]].
		%29 citations counted in INSPIRE as of 16 Jul 2023
		
		
		
		
		%\cite{Vagnozzi:2020gtf}
		\bibitem{Vagnozzi:2020gtf}
		S.~Vagnozzi,
		%``Implications of the NANOGrav results for inflation,''
		Mon. Not. Roy. Astron. Soc. \textbf{502}, no.1, L11-L15 (2021)
		%doi:10.1093/mnrasl/slaa203
		[arXiv:2009.13432 [astro-ph.CO]].
		%108 citations counted in INSPIRE as of 16 Jul 2023
		
		
	
		
		
	
		
		
		
		%\cite{Buchmuller:2021dtt}
		\bibitem{Buchmuller:2021dtt}
		W.~Buchmuller,
		%``Metastable strings and dumbbells in supersymmetric hybrid inflation,''
		JHEP \textbf{04}, 168 (2021)
	%	doi:10.1007/JHEP04(2021)168
		[arXiv:2102.08923 [hep-ph]].
		%12 citations counted in INSPIRE as of 05 Aug 2023
		
		%\cite{Buchmuller:2021mbb}
		\bibitem{Buchmuller:2021mbb}
		W.~Buchmuller, V.~Domcke and K.~Schmitz,
		%``Stochastic gravitational-wave background from metastable cosmic strings,''
		JCAP \textbf{12}, no.12, 006 (2021)
		%doi:10.1088/1475-7516/2021/12/006
		[arXiv:2107.04578 [hep-ph]].
		%44 citations counted in INSPIRE as of 05 Aug 2023
		
		%\cite{Masoud:2021prr}
		\bibitem{Masoud:2021prr}
		M.~A.~Masoud, M.~U.~Rehman and Q.~Shafi,
		%``Sneutrino tribrid inflation, metastable cosmic strings and gravitational waves,''
		JCAP \textbf{11}, 022 (2021)
		%doi:10.1088/1475-7516/2021/11/022
		[arXiv:2107.09689 [hep-ph]].
		%18 citations counted in INSPIRE as of 05 Aug 2023
		
		
		
		
		%\cite{}
	
		
		%\cite{Jeannerot:2002wt}
		\bibitem{Jeannerot:2002wt}
		R.~Jeannerot, S.~Khalil and G.~Lazarides,
		%``New shifted hybrid inflation,''
		JHEP \textbf{07}, 069 (2002)
	%	doi:10.1088/1126-6708/2002/07/069
		[arXiv:hep-ph/0207244 [hep-ph]].
		%47 citations counted in INSPIRE as of 16 Jul 2023

  
%\cite{Ahmed:2022wed}
\bibitem{Ahmed:2022wed}
W.~Ahmed, M.~Moosa, S.~Munir and U.~Zubair,
%``Observable r, gravitino dark matter, and non-thermal leptogenesis in no-scale supergravity,''
JHEP \textbf{05}, 011 (2023)
%doi:10.1007/JHEP05(2023)011
[arXiv:2208.11888 [hep-ph]].
%1 citations counted in INSPIRE as of 17 Aug 2023
  
		
		
		%\cite{Khalil:2010cp}
		\bibitem{Khalil:2010cp}
		S.~Khalil, M.~U.~Rehman, Q.~Shafi and E.~A.~Zaakouk,
		%``Inflation in Supersymmetric SU(5),''
		Phys. Rev. D \textbf{83}, 063522 (2011)
		%doi:10.1103/PhysRevD.83.063522
		[arXiv:1010.3657 [hep-ph]].
		%23 citations counted in INSPIRE as of 16 Jul 2023

  %\cite{Ahmed:2022vlc}
\bibitem{Ahmed:2022vlc}
W.~Ahmed, A.~Karozas, G.~K.~Leontaris and U.~Zubair,
%``Smooth hybrid inflation with low reheat temperature and observable gravity waves in SU(5) \texttimes{} U(1)$_{χ}$ super-GUT,''
JCAP \textbf{06}, no.06, 027 (2022)
%doi:10.1088/1475-7516/2022/06/027
[arXiv:2201.12789 [hep-ph]].
%6 citations counted in INSPIRE as of 13 Aug 2023


  %\cite{Rehman:2014rpa}
\bibitem{Rehman:2014rpa}
M.~U.~Rehman and U.~Zubair,
%``Simplified Smooth Hybrid Inflation in Supersymmetric SU(5),''
Phys. Rev. D \textbf{91}, 103523 (2015)
%doi:10.1103/PhysRevD.91.103523
[arXiv:1412.7619 [hep-ph]].
%8 citations counted in INSPIRE as of 13 Aug 2023


%\cite{Arai:2011nq}
\bibitem{Arai:2011nq}
M.~Arai, S.~Kawai and N.~Okada,
%``Higgs inflation in minimal supersymmetric SU(5) GUT,''
Phys. Rev. D \textbf{84}, 123515 (2011)
%doi:10.1103/PhysRevD.84.123515
[arXiv:1107.4767 [hep-ph]].
%44 citations counted in INSPIRE as of 13 Aug 2023

\bibitem{Masoud:2019cen}
M.~A.~Masoud, M.~U.~Rehman and M.~M.~A.~Abid,
%``Nonminimal inflation in supersymmetric GUTs with $U(1)_R \times Z_n$ symmetry,''
Int. J. Mod. Phys. D \textbf{28}, no.16, 2040015 (2019)
%doi:10.1142/S0218271820400155
[arXiv:1910.10519 [hep-ph]].


%\cite{Rehman:2018gnr}
\bibitem{Rehman:2018gnr}
M.~U.~Rehman, M.~M.~A.~Abid and A.~Ejaz,
%``New Inflation in Supersymmetric SU(5) and Flipped SU(5) GUT Models,''
JCAP \textbf{11}, 019 (2020)
%doi:10.1088/1475-7516/2020/11/019
[arXiv:1804.07619 [hep-ph]].
%6 citations counted in INSPIRE as of 14 Aug 2023

%\cite{Kibble:1982ae}
\bibitem{Kibble:1982ae}
T.~W.~B.~Kibble, G.~Lazarides and Q.~Shafi,
%``Strings in SO(10),''
Phys. Lett. B \textbf{113}, 237-239 (1982)
%doi:10.1016/0370-2693(82)90829-2
%237 citations counted in INSPIRE as of 13 Aug 2023


%\cite{Nelson:1993nf}
\bibitem{Nelson:1993nf}
A.~E.~Nelson and N.~Seiberg,
%``R symmetry breaking versus supersymmetry breaking,''
Nucl. Phys. B \textbf{416}, 46-62 (1994)
%doi:10.1016/0550-3213(94)90577-0
[arXiv:hep-ph/9309299 [hep-ph]].
%386 citations counted in INSPIRE as of 13 Aug 2023


%\cite{Civiletti:2013cra}
\bibitem{Civiletti:2013cra}
M.~Civiletti, M.~Ur Rehman, E.~Sabo, Q.~Shafi and J.~Wickman,
%``R-symmetry breaking in supersymmetric hybrid inflation,''
Phys. Rev. D \textbf{88}, no.10, 103514 (2013)
%doi:10.1103/PhysRevD.88.103514
[arXiv:1303.3602 [hep-ph]].
%13 citations counted in INSPIRE as of 13 Aug 2023


%\cite{Dvali:1997uq}
\bibitem{Dvali:1997uq}
G.~R.~Dvali, G.~Lazarides and Q.~Shafi,
%``Mu problem and hybrid inflation in supersymmetric SU(2)-L x SU(2)-R x U(1)-(B-L),''
Phys. Lett. B \textbf{424}, 259-264 (1998)
%doi:10.1016/S0370-2693(98)00145-2
[arXiv:hep-ph/9710314 [hep-ph]].
%131 citations counted in INSPIRE as of 15 Aug 2023

%\cite{Albright:1997xw}
\bibitem{Albright:1997xw}
C.~H.~Albright and S.~M.~Barr,
%``Fermion masses in SO(10) with a single adjoint Higgs field,''
Phys. Rev. D \textbf{58}, 013002 (1998)
%doi:10.1103/PhysRevD.58.013002
[arXiv:hep-ph/9712488 [hep-ph]].
%130 citations counted in INSPIRE as of 17 Aug 2023

%\cite{Albright:1998sy}
\bibitem{Albright:1998sy}
C.~H.~Albright, K.~S.~Babu and S.~M.~Barr,
%``Implications of a minimal SO(10) Higgs structure,''
Nucl. Phys. B Proc. Suppl. \textbf{77}, 308-312 (1999)
%doi:10.1016/S0920-5632(99)00433-8
[arXiv:hep-ph/9805266 [hep-ph]].
%16 citations counted in INSPIRE as of 17 Aug 2023


%\cite{Kyae:2005vg}
\bibitem{Kyae:2005vg}
B.~Kyae and Q.~Shafi,
%``Inflation with realistic supersymmetric SO(10),''
Phys. Rev. D \textbf{72}, 063515 (2005)
%doi:10.1103/PhysRevD.72.063515
[arXiv:hep-ph/0504044 [hep-ph]].
%39 citations counted in INSPIRE as of 17 Aug 2023


  %\cite{Barr:2005xya}
\bibitem{Barr:2005xya}
S.~M.~Barr, B.~Kyae and Q.~Shafi,
%``Flat-directions in grand unification with U(1)(R) symmetry,''
[arXiv:hep-ph/0511097 [hep-ph]].
%4 citations counted in INSPIRE as of 15 Aug 2023

%\cite{Fallbacher:2011xg}
\bibitem{Fallbacher:2011xg}
M.~Fallbacher, M.~Ratz and P.~K.~S.~Vaudrevange,
%``No-go theorems for R symmetries in four-dimensional GUTs,''
Phys. Lett. B \textbf{705}, 503-506 (2011)
%doi:10.1016/j.physletb.2011.10.063
[arXiv:1109.4797 [hep-ph]].
%34 citations counted in INSPIRE as of 15 Aug 2023


%\cite{Masoud:2019gxx}
\bibitem{Masoud:2019gxx}
M.~A.~Masoud, M.~U.~Rehman and Q.~Shafi,
%``Pseudosmooth Tribrid Inflation in $SU(5)$,''
JCAP \textbf{04}, 041 (2020)
%doi:10.1088/1475-7516/2020/04/041
[arXiv:1910.07554 [hep-ph]].
%6 citations counted in INSPIRE as of 17 Aug 2023

 
		%\cite{Ahmed:2022thr}
		\bibitem{Ahmed:2022thr}
		W.~Ahmed and U.~Zubair,
		%``Radiative symmetry breaking, cosmic strings and observable gravity waves in \ensuremath{\mathsf{U}}(1)$_{��}$ symmetric \ensuremath{\mathsf{S}}\ensuremath{\mathsf{U}}(5) \texttimes{} \ensuremath{\mathsf{U}}(1)$_{χ}$,''
		JCAP \textbf{01}, 019 (2023)
		%doi:10.1088/1475-7516/2023/01/019
		[arXiv:2210.13059 [hep-ph]].
		%0 citations counted in INSPIRE as of 05 Aug 2023
		


   
		
  %\cite{Ahmed:2021dvo}
\bibitem{Ahmed:2021dvo}
W.~Ahmed, A.~Karozas and G.~K.~Leontaris,
%``Gravitino dark matter, nonthermal leptogenesis, and low reheating temperature in no-scale Higgs inflation,''
Phys. Rev. D \textbf{104}, no.5, 055025 (2021)
%doi:10.1103/PhysRevD.104.055025
[arXiv:2104.04328 [hep-ph]].
%6 citations counted in INSPIRE as of 17 Aug 2023


  %\cite{Ahmed:2018jlv}
\bibitem{Ahmed:2018jlv}
W.~Ahmed and A.~Karozas,
%``Inflation from a no-scale supersymmetric $SU(4)_{c}\times{SU(2)_{L}\times{SU(2)_{R}}}$ model,''
Phys. Rev. D \textbf{98}, no.2, 023538 (2018)
%doi:10.1103/PhysRevD.98.023538
[arXiv:1804.04822 [hep-ph]].
%18 citations counted in INSPIRE as of 17 Aug 2023

  %\cite{Cicoli:2013rwa}
		\bibitem{Cicoli:2013rwa}
		M.~Cicoli, S.~de Alwis and A.~Westphal,
		%``Heterotic Moduli Stabilisation,''
		JHEP \textbf{10}, 199 (2013)   
	%	doi:10.1007/JHEP10(2013)199
		[arXiv:1304.1809 [hep-th]].
		%61 citations counted in INSPIRE as of 05 Aug 2023
		
		
		%\cite{Ellis:2013nxa}
		\bibitem{Ellis:2013nxa}
		J.~Ellis, D.~V.~Nanopoulos and K.~A.~Olive,
		%``Starobinsky-like Inflationary Models as Avatars of No-Scale Supergravity,''
		JCAP \textbf{10}, 009 (2013)
	%	doi:10.1088/1475-7516/2013/10/009
		[arXiv:1307.3537 [hep-th]].
		%236 citations counted in INSPIRE as of 05 Aug 2023
		

		
		
		%\cite{Lee:2010hj}
		\bibitem{Lee:2010hj}
		H.~M.~Lee,
		%``Chaotic inflation in Jordan frame supergravity,''
		JCAP \textbf{08}, 003 (2010)
	%	doi:10.1088/1475-7516/2010/08/003
		[arXiv:1005.2735 [hep-ph]].
		%92 citations counted in INSPIRE as of 05 Aug 2023
		
		
		
		%\cite{Kolb:1990vq}
		\bibitem{Kolb:1990vq}
		E.~W.~Kolb and M.~S.~Turner,
		%``The Early Universe,''
		Front. Phys. \textbf{69}, 1-547 (1990)
	%	doi:10.1201/9780429492860
		%2154 citations counted in INSPIRE as of 05 Aug 2023
  
		
		  %\cite{Pallis:2018ver}
\bibitem{Pallis:2018ver}
C.~Pallis and Q.~Shafi,
%``Induced-Gravity GUT-Scale Higgs Inflation in Supergravity,''
Eur. Phys. J. C \textbf{78}, no.6, 523 (2018)
%doi:10.1140/epjc/s10052-018-5980-0
[arXiv:1803.00349 [hep-ph]].
%13 citations counted in INSPIRE as of 13 Aug 2023 


%\cite{Pallis:2018acu}
\bibitem{Pallis:2018acu}
C.~Pallis,
%``$B-L$ Higgs Inflation in Supergravity With Several Consequences,''
PoS \textbf{CORFU2017}, 086 (2018)
%doi:10.22323/1.318.0086
[arXiv:1804.07038 [hep-ph]].
%4 citations counted in INSPIRE as of 13 Aug 2023



%\cite{Endo:2006qk}
\bibitem{Endo:2006qk}
M.~Endo, M.~Kawasaki, F.~Takahashi and T.~T.~Yanagida,
``Inflaton decay through supergravity effects,''
Phys. Lett. B \textbf{642}, 518-524 (2006)
%doi:10.1016/j.physletb.2006.09.044
[arXiv:hep-ph/0607170 [hep-ph]].





%\cite{Endo:2007sz}
\bibitem{Endo:2007sz}
M.~Endo, F.~Takahashi and T.~T.~Yanagida,
``Inflaton Decay in Supergravity,''
Phys. Rev. D \textbf{76}, 083509 (2007)
%doi:10.1103/PhysRevD.76.083509
[arXiv:0706.0986 [hep-ph]].

%\cite{Pallis:2011gr}
\bibitem{Pallis:2011gr}
C.~Pallis and N.~Toumbas,
``Non-Minimal Higgs Inflation and non-Thermal Leptogenesis in A Supersymmetric Pati-Salam Model,''
JCAP \textbf{12}, 002 (2011)
%doi:10.1088/1475-7516/2011/12/002
[arXiv:1108.1771 [hep-ph]].

%\cite{Ellis:1984eq}
\bibitem{Ellis:1984eq}
J.~R.~Ellis, J.~E.~Kim and D.~V.~Nanopoulos,
%``Cosmological Gravitino Regeneration and Decay,''
Phys. Lett. B \textbf{145}, 181-186 (1984)
doi:10.1016/0370-2693(84)90334-4
%906 citations counted in INSPIRE as of 17 Aug 2023

%\cite{Khlopov:1984pf}
\bibitem{Khlopov:1984pf}
M.~Y.~Khlopov and A.~D.~Linde,
%``Is It Easy to Save the Gravitino?,''
Phys. Lett. B \textbf{138}, 265-268 (1984)
%doi:10.1016/0370-2693(84)91656-3
%867 citations counted in INSPIRE as of 17 Aug 2023



%\cite{LiteBIRD:2020khw}
\bibitem{LiteBIRD:2020khw}
M.~Hazumi \textit{et al.} [LiteBIRD],
%``LiteBIRD: JAXA's new strategic L-class mission for all-sky surveys of cosmic microwave background polarization,''
Proc. SPIE Int. Soc. Opt. Eng. \textbf{11443}, 114432F (2020)
%doi:10.1117/12.2563050
[arXiv:2101.12449 [astro-ph.IM]].
%99 citations counted in INSPIRE as of 14 Aug 2023




%\cite{SimonsObservatory:2018koc}
\bibitem{SimonsObservatory:2018koc}
P.~Ade \textit{et al.} [Simons Observatory],
%``The Simons Observatory: Science goals and forecasts,''
JCAP \textbf{02}, 056 (2019)
%doi:10.1088/1475-7516/2019/02/056
[arXiv:1808.07445 [astro-ph.CO]].
%929 citations counted in INSPIRE as of 14 Aug 2023


 %\cite{Andre:2013afa}
 \bibitem{Andre:2013afa}
 P.~Andre \textit{et al.} [PRISM],
 %``PRISM (Polarized Radiation Imaging and Spectroscopy Mission): A White Paper on the Ultimate Polarimetric Spectro-Imaging of the Microwave and Far-Infrared Sky,''
 [arXiv:1306.2259 [astro-ph.CO]].
 %154 citations counted in INSPIRE as of 06 Sep 2020

%\cite{CMB-S4:2020lpa}
\bibitem{CMB-S4:2020lpa}
K.~Abazajian \textit{et al.} [CMB-S4],
%``CMB-S4: Forecasting Constraints on Primordial Gravitational Waves,''
Astrophys. J. \textbf{926}, no.1, 54 (2022)
%doi:10.3847/1538-4357/ac1596
[arXiv:2008.12619 [astro-ph.CO]].
%147 citations counted in INSPIRE as of 14 Aug 2023

%\cite{Sehgal:2019ewc}
\bibitem{Sehgal:2019ewc}
N.~Sehgal, S.~Aiola, Y.~Akrami, K.~Basu, M.~Boylan-Kolchin, S.~Bryan, S.~Clesse, F.~Y.~Cyr-Racine, L.~Di Mascolo and S.~Dicker, \textit{et al.}
%``CMB-HD: An Ultra-Deep, High-Resolution Millimeter-Wave Survey Over Half the Sky,''
[arXiv:1906.10134 [astro-ph.CO]].
%83 citations counted in INSPIRE as of 14 Aug 2023


  
  %\cite{Finelli:2016cyd}
  \bibitem{Finelli:2016cyd}
  F.~Finelli \textit{et al.} [CORE],
  %``Exploring cosmic origins with CORE: Inflation,''
  JCAP \textbf{04}, 016 (2018)
  %doi:10.1088/1475-7516/2018/04/016
  [arXiv:1612.08270 [astro-ph.CO]].
  %120 citations counted in INSPIRE as of 06 Sep 2020

%\cite{Kogut:2011xw}
 \bibitem{Kogut:2011xw}
 A.~Kogut, D.~J.~Fixsen, D.~T.~Chuss, J.~Dotson, E.~Dwek, M.~Halpern, G.~F.~Hinshaw, S.~M.~Meyer, S.~H.~Moseley, M.~D.~Seiffert, D.~N.~Spergel and E.~J.~Wollack,
 %``The Primordial Inflation Explorer (PIXIE): A Nulling Polarimeter for Cosmic Microwave Background Observations,''
 JCAP \textbf{07}, 025 (2011)
 %doi:10.1088/1475-7516/2011/07/025
 [arXiv:1105.2044 [astro-ph.CO]].
 %437 citations counted in INSPIRE as of 06 Sep 2020


 






%\cite{Planck:2018jri}
\bibitem{Planck:2018jri}
Y.~Akrami \textit{et al.} [Planck],
%``Planck 2018 results. X. Constraints on inflation,''
Astron. Astrophys. \textbf{641}, A10 (2020)
%doi:10.1051/0004-6361/201833887
[arXiv:1807.06211 [astro-ph.CO]].
%2607 citations counted in INSPIRE as of 13 Aug 2023


			%\cite{Planck:2018vyg}
\bibitem{Planck:2018vyg}
N.~Aghanim \textit{et al.} [Planck],
%``Planck 2018 results. VI. Cosmological parameters,''
Astron. Astrophys. \textbf{641}, A6 (2020)
[erratum: Astron. Astrophys. \textbf{652}, C4 (2021)]
%doi:10.1051/0004-6361/201833910
[arXiv:1807.06209 [astro-ph.CO]].
%11290 citations counted in INSPIRE as of 13 Aug 2023


	
		%\cite{Blanco-Pillado:2017oxo}
		\bibitem{Blanco-Pillado:2017oxo}
		J.~J.~Blanco-Pillado and K.~D.~Olum,
		%``Stochastic gravitational wave background from smoothed cosmic string loops,''
		Phys. Rev. D \textbf{96}, no.10, 104046 (2017)
	%	doi:10.1103/PhysRevD.96.104046
		[arXiv:1709.02693 [astro-ph.CO]].
		%151 citations counted in INSPIRE as of 06 Aug 2023
		



		
		
	
		
		%\cite{Auclair:2019wcv}
		\bibitem{Auclair:2019wcv}
		P.~Auclair, J.~J.~Blanco-Pillado, D.~G.~Figueroa, A.~C.~Jenkins, M.~Lewicki, M.~Sakellariadou, S.~Sanidas, L.~Sousa, D.~A.~Steer and J.~M.~Wachter, \textit{et al.}
		%``Probing the gravitational wave background from cosmic strings with LISA,''
		JCAP \textbf{04}, 034 (2020)
	%	doi:10.1088/1475-7516/2020/04/034
		[arXiv:1909.00819 [astro-ph.CO]].
		%195 citations counted in INSPIRE as of 06 Aug 2023
		

		%\cite{Leblond:2009fq}
		\bibitem{Leblond:2009fq}
		L.~Leblond, B.~Shlaer and X.~Siemens,
		%``Gravitational Waves from Broken Cosmic Strings: The Bursts and the Beads,''
		Phys. Rev. D \textbf{79}, 123519 (2009)
		%doi:10.1103/PhysRevD.79.123519
		[arXiv:0903.4686 [astro-ph.CO]].
		%52 citations counted in INSPIRE as of 09 Feb 2022
		
		%\cite{Monin:2008mp}
		\bibitem{Monin:2008mp}
		A.~Monin and M.~B.~Voloshin,
		%``The Spontaneous breaking of a metastable string,''
		Phys. Rev. D \textbf{78}, 065048 (2008)
		%doi:10.1103/PhysRevD.78.065048
		[arXiv:0808.1693 [hep-th]].
		%21 citations counted in INSPIRE as of 09 Feb 2022
		
		\bibitem{KaiSchmitz}
		Wilfried Buchmuller, Valerie Domcke and Kai Schmitz,
		Phys. Lett. B 811 (2020) 135914 
		[arXiv:2009.10649 [astro-ph.CO]];
		JCAP 12 (2021) 12, 006,
		[arXiv:2107.04578 [hep-ph]].

  \bibitem{LIGOScientific:2019vic}
		B.~P.~Abbott \textit{et al.} [LIGO Scientific and Virgo],
		%``Search for the isotropic stochastic background using data from Advanced LIGO\textquoteright{}s second observing run,''
		Phys. Rev. D \textbf{100}, no.6, 061101 (2019)
		%doi:10.1103/PhysRevD.100.061101
		[arXiv:1903.02886 [gr-qc]].
		%192 citations counted in INSPIRE as of 09 Feb 2022
  
		
	
		
		
		
	



		%\cite{Ferdman:2010xq}
		\bibitem{Ferdman:2010xq}
		R.~D.~Ferdman, R.~van Haasteren, C.~G.~Bassa, M.~Burgay, I.~Cognard, A.~Corongiu, N.~D'Amico, G.~Desvignes, J.~W.~T.~Hessels and G.~H.~Janssen, \textit{et al.}
		%``The European Pulsar Timing Array: current efforts and a LEAP toward the future,''
		Class. Quant. Grav. \textbf{27}, 084014 (2010)
		%doi:10.1088/0264-9381/27/8/084014
		[arXiv:1003.3405 [astro-ph.HE]].
		%85 citations counted in INSPIRE as of 10 Feb 2022



	%\cite{Smits:2008cf}
		\bibitem{Smits:2008cf}
		R.~Smits, M.~Kramer, B.~Stappers, D.~R.~Lorimer, J.~Cordes and A.~Faulkner,
		%``Pulsar searches and timing with the square kilometre array,''
		Astron. Astrophys. \textbf{493} (2009), 1161-1170
		%doi:10.1051/0004-6361:200810383
		[arXiv:0811.0211 [astro-ph]].
		%128 citations counted in INSPIRE as of 10 Feb 2022
  	
		
		%\cite{LISA:2017pwj}
		\bibitem{LISA:2017pwj}
		P.~Amaro-Seoane \textit{et al.} [LISA],
		%``Laser Interferometer Space Antenna,''
		[arXiv:1702.00786 [astro-ph.IM]].
		%1494 citations counted in INSPIRE as of 09 Feb 2022
		
		
		
		
		

		
		
	
		
		
		
		%\cite{Hu:2017mde}
		\bibitem{Hu:2017mde}
		W.~R.~Hu and Y.~L.~Wu,
		%``The Taiji Program in Space for gravitational wave physics and the nature of gravity,''
		Natl. Sci. Rev. \textbf{4}, no.5, 685-686 (2017)
		%doi:10.1093/nsr/nwx116
		%185 citations counted in INSPIRE as of 10 Feb 2022
		
		
		%\cite{TianQin:2015yph}
		\bibitem{TianQin:2015yph}
		J.~Luo \textit{et al.} [TianQin],
		%``TianQin: a space-borne gravitational wave detector,''
		Class. Quant. Grav. \textbf{33} (2016) no.3, 035010
		%doi:10.1088/0264-9381/33/3/035010
		[arXiv:1512.02076 [astro-ph.IM]].
		%484 citations counted in INSPIRE as of 10 Feb 2022
		
		
		
		%\cite{Corbin:2005ny}
		\bibitem{Corbin:2005ny}
		V.~Corbin and N.~J.~Cornish,
		%``Detecting the cosmic gravitational wave background with the big bang observer,''
		Class. Quant. Grav. \textbf{23}, 2435-2446 (2006)
		%doi:10.1088/0264-9381/23/7/014
		[arXiv:gr-qc/0512039 [gr-qc]].
		%203 citations counted in INSPIRE as of 11 Feb 2022
		
		%\cite{Seto:2001qf}
		\bibitem{Seto:2001qf}
		N.~Seto, S.~Kawamura and T.~Nakamura,
		%``Possibility of direct measurement of the acceleration of the universe using 0.1-Hz band laser interferometer gravitational wave antenna in space,''
		Phys. Rev. Lett. \textbf{87}, 221103 (2001)
		%doi:10.1103/PhysRevLett.87.221103
		[arXiv:astro-ph/0108011 [astro-ph]].
		%536 citations counted in INSPIRE as of 11 Feb 2022
		
		
		
		
		
		
		
		%\cite{Punturo:2010zz}
		\bibitem{Punturo:2010zz}
		M.~Punturo, M.~Abernathy, F.~Acernese, B.~Allen, N.~Andersson, K.~Arun, F.~Barone, B.~Barr, M.~Barsuglia and M.~Beker, \textit{et al.}
		%``The Einstein Telescope: A third-generation gravitational wave observatory,''
		Class. Quant. Grav. \textbf{27} (2010), 194002
		%doi:10.1088/0264-9381/27/19/194002
		%886 citations counted in INSPIRE as of 10 Feb 2022
		
		%\cite{LIGOScientific:2016wof}
		\bibitem{LIGOScientific:2016wof}
		B.~P.~Abbott \textit{et al.} [LIGO Scientific],
		%``Exploring the Sensitivity of Next Generation Gravitational Wave Detectors,''
		Class. Quant. Grav. \textbf{34}, no.4, 044001 (2017)
		%doi:10.1088/1361-6382/aa51f4
		[arXiv:1607.08697 [astro-ph.IM]].
		%632 citations counted in INSPIRE as of 11 Feb 2022
		
		%\cite{AEDGE:2019nxb}
		\bibitem{AEDGE:2019nxb}
		Y.~A.~El-Neaj \textit{et al.} [AEDGE],
		%``AEDGE: Atomic Experiment for Dark Matter and Gravity Exploration in Space,''
		EPJ Quant. Technol. \textbf{7}, 6 (2020)
		%doi:10.1140/epjqt/s40507-020-0080-0
		[arXiv:1908.00802 [gr-qc]].
		%101 citations counted in INSPIRE as of 11 Feb 2022
		
		
		
		
		
		
		
		%%%%%%%%%%%%%%%%%%%%%%%%%%%%%%%
	\end{thebibliography}
\end{document}